\documentclass{aa}
\usepackage{graphics}
\usepackage{txfonts,natbib}
\bibpunct{(}{)}{;}{a}{}{,}

\def\bx{{\bf x}}
\def\by{{\bf y}}
\def\bk{{\bf k}}
\def\bq{{\bf q}}
\def\bp{{\bf p}}
\def\bu{{\bf u}}
\def\ki{{\kappa}_{\rm int}}
\def\km{{\kappa}_{\rm mac}}

\begin{document}

\title{Numerical study of the cosmological velocity field \\ as a
  function of density}

\author{Alvaro Dom\'\i nguez\inst{1,2} \and Adrian L. Melott\inst{3}}

\institute{Max--Planck--Institut f\"ur Metallforschung,
  Heisenbergstr.~3, D--70569 Stuttgart, Germany \\
  \email{alvaro@fluids.mpi-stuttgart.mpg.de} 
  \and
  Theoretische Physik, Ludwig--Maximilians--Universit\"at,
  Theresienstr. 37, D--80333 M\"unchen, Germany
  \and
  Department of Physics and Astronomy, University of Kansas, Lawrence, Kansas 66045, USA}

\date{March 1, 2004}

\titlerunning{Cosmological velocity field as a function of
  density}

\authorrunning{Dom\'\i nguez \and Melott}

\abstract{
  We report on a new study of the velocity distribution in
  N-body simulations.
  We investigate the center--of--mass and internal kinetic energies of
  coarsening cells as a function of time, cell size and cell mass. By
  using self--similar cosmological models, we are able 
  to derive theoretical predictions for comparison and
  to assess 
  the influence of finite--size and resolution effects.
  The most interesting
  result is the discovery of a polytropic--like relationship between
  the average velocity dispersion (internal kinetic energy) and the mass density in an intermediate
  range of densities, $\ki \propto \varrho^{2-\eta}$. The exponent
  $\eta$ measures the deviations from the virial prediction,
  $\eta_{\rm virial}=0$. For self--similar models, $\eta$ depends only on
  the spectral index of the initial power spectrum. We also study CDM models and confirm a
  previous result that the same polytropic--like dependence exists, 
  with a time and coarsening length dependent $\eta$.
  The dependence $\ki(\varrho)$ is an important input for a
  recently proposed theoretical model of cosmological structure
  formation which improves over the standard dust model (pressureless fluid) by regularizing
  the density singularities.

}

\maketitle

\section{Introduction}

The use of N-body simulations has proved a useful tool in the
investigation of both the cosmological structure formation and the
evolution by self--gravity. The main interest has been
concentrated in properties of the spatial distribution of matter (mass
correlations, void distribution, morphological features...),
while the kinetic properties have received comparatively little
attention, without doubt due to the larger difficulties to obtain
reliable kinetic measurements from real data with which to compare \citep{WFCG02}.

The kinetic measurements addressed till
now with N-body simulations have pertained the quasilinear velocity field 
(see e.g.~the reviews by \citealt{Deke94} and \citealt{BCGS02}; an application closely related to the present work is \citealt{SeSu01}),
the pairwise relative velocity (\citealt{Peeb80}; for recent
applications, see e.g.~\citealt{SOC98,FJFD03}),
and the velocity dispersion of halos (e.g.~\citealt{KnMu99} in connection with the present work).
Our work
addresses the center--of--mass velocity of coarsening cells ({\it
  macroscopic kinetic energy}), as well as the velocity dispersion of
the particles inside the cell ({\it internal kinetic energy}). The
coarsening cells are randomly centered and of variable size (probing
both the linear and the nonlinear regimes); in this way our analysis
does not suffer the arbitrariness intrinsic to the definition of
clusters and halos \citep[ and refs.~therein]{KlHo97},
and essentially all the simulation particles are employed in the
determination of the quantities.
We are aware of two works where a similar analysis of
N-body simulations has been performed \citep{KSS97,NOC01},
motivated differently than ours.
The connection to our work is explained in Sec.~\ref{sec:discussion} in detail.

The present study focuses on the dependence of the cell kinetic
energies on the cell mass density. The main motivation is the
application to models of cosmological structure formation by
self--gravity. The most widely used theoretical model is the dust model
(pressureless fluid) \citep{Peeb80,Padm95}, which has been studied
intensively \citep[see e.g.~the reviews][]{SaCo95,BCGS02} but has the
shortcoming of producing singularities. Some recent works
\citep{BuDo98,AdBu99,BDP99,MTM99,Domi00,MoTa01,Domi02,TSMM02} have
proposed a novel approach. One of its features is the ability to
derive adhesion--like models \citep{KoSh88,GSS89,KPSM92,MSW94,SSMP95},
and to offer a possible explanation of the physical origin for the
``adhesive'' behavior which regularizes the mass density singularities
of the dust model. In these improved models, the internal kinetic
energy brings about the ``adhesive'' behavior provided it can be
approximated as a function of density and/or the gradients of the
velocity field.

Another goal of this work is to confirm the results by
\citet{Domi99,Domi03}, where a polytropic--like dependence between
internal kinetic energy and mass density is found in AP$^3$M
simulations of CDM models. We indeed corroborate this finding in PM
simulations of CDM models and also of self--similar models. The latter
are particularly amenable to a theoretical analysis and allow the
identification of the influence of finite--size and resolution effects
in the measurements. We conclude that the polytropic--like dependence
is unlikely to be an artifact of the simulations.

The paper is organized as follows: in Sec.~\ref{sec:theory} we work
out the theoretical predictions for the density dependence of the
macroscopic and internal kinetic energies.
In Sec.~\ref{sec:method} we describe the simulations and the method
how we measure the kinetic energies. In Sec.~\ref{sec:result} we
present the results of the analysis. Sec.~\ref{sec:discussion}
contains a discussion of the results and the conclusions.

\section{Theoretical background}
\label{sec:theory}

Let $a(t)$ denote the cosmological expansion factor, $m$ the mass of a particle,
and ${\bf x}_\alpha$ and ${\bf u}_\alpha$ the comoving position and
peculiar velocity, respectively, of the $\alpha$-th particle.
$W(\cdot)$ is a (normalized) smoothing window. Then, given a comoving
smoothing scale $L$, the coarse--grained mass density field, velocity
field and density of {\em internal} peculiar kinetic energy are
defined respectively as follows:
\begin{displaymath}
  \varrho ({\bf x}, t; L) = \frac{m}{[a(t) L]^3} \sum_{\alpha=1}^{N} \, 
  W \left({{\bf x}-{\bf x}_\alpha (t) \over L} \right) , 
\end{displaymath}
\begin{equation}
  \label{eq:fields}
  \varrho {\bf u} ({\bf x}, t; L) = \frac{m}{[a(t) L]^3} 
  \sum_{\alpha=1}^{N} \, {\bf u}_\alpha(t) \, 
  W \left({{\bf x}-{\bf x}_\alpha (t) \over L} \right) ,
\end{equation}
\begin{displaymath}
  K_{\rm int} ({\bf x}, t; L) = \frac{m}{2 [a(t) L]^3} \sum_{\alpha=1}^{N}
  |{\bf u}_{\alpha}(t) - {\bf u}({\bf x}, t; L)|^2
  W \left({{\bf x}-{\bf x}_\alpha (t) \over L} \right) .
\end{displaymath}
The density of {\em macroscopic}\footnote{Borrowing the nomenclature from statistical physics.} peculiar kinetic energy is defined as
$K_{\rm mac} = (1/2) \varrho |{\bf u}|^2$, so that one can write
$K_{tot}=K_{\rm int}+K_{\rm mac}$, where $K_{tot}$ is the density of {\em
  total} peculiar kinetic energy: this is given by the same expression
as $K_{\rm int}$ above but dropping ${\bf u}$. Whether $K_{\rm mac}$ or
$K_{\rm int}$ dominates the contribution to $K_{tot}$ means respectively
that the particle velocities ${\bf u}_\alpha$ are controlled by the
center--of--mass motion of the coarsening cell as a whole or by
``internal'' motions within the cell (the ratio $K_{\rm mac}/K_{\rm int}$ was introduced as the cosmic Mach number by \citet{OsSu90}). Indeed, $K_{\rm int}$ is a much
better probe of the dynamics at the small scales than $K_{\rm mac}$,
because in the latter there can be extensive cancellations in the
vectorial sum defining ${\bf u}$. 

The
purpose of the present study is the
relationship between $K_{\rm int}$ and the field $\varrho$.
Our main interest is
the density dependence of $\langle K_{\rm int} \rangle_\varrho$, that is, the average of $K_{\rm int}$
conditioned to a given value $\varrho$ of the coarse--grained mass density,
\begin{equation}
  \label{eq:eqstate}
  \ki (\varrho, t; L) := \langle K_{\rm int} \rangle_\varrho = 
  \frac{\langle K_{\rm int} ({\bf x}, t; L) 
    \delta^{(3)} (\varrho({\bf x}, t; L)-\varrho) \rangle}
  {\langle \delta^{(3)} (\varrho({\bf x}, t; L)-\varrho) \rangle},
\end{equation}
where $\langle \cdot \rangle$ denotes ensemble average, which by
translational invariance must be ${\bf x}$--independent, and $\delta^{(3)}(\cdot)$ is Dirac's delta function. In the same
way is defined the conditioned average $\km:=\langle
K_{\rm mac} \rangle_\varrho$.
Since $\ki$ receives contributions from
the highly non-linear regime, the density dependence cannot be
computed analytically in general.  Nevertheless, a lot can be learned
by way of suitable approximations, whose validity will be checked by
comparing with simulations.

To simplify the theoretical discussion, we consider a self--similar
cosmological model: an Einstein--de Sitter background and an initial
Gaussian distributed density field with power spectrum
$P(k, t_{initial}) = A k^n$, with the bounds $n>-3$ (so that density
fluctuations do not receive a divergent contribution from $k
\rightarrow 0$) and $n<4$ (imposed by the unavoidable graininess due
to the point particles) \citep{Peeb80,Padm95}. The conclusions should
apply qualitatively unaltered to a more realistic case. Let $\sigma^2
(L, t)=\langle \delta^2 ({\bf x}, t; L) \rangle$ denote the variance
of the density contrast smoothed on the scale $L$ ($\delta :=
(\varrho/\varrho_b) -1$ is the density contrast, with $\varrho_b :=
\langle \varrho \rangle$ the background density),
\begin{displaymath}
  \sigma^2 (L, t) = \int \!\! \frac{d{\bf k}}{(2 \pi)^3} \; P(k, t) 
  |\tilde{W} (L \bk)|^2 .
\end{displaymath}
(A tilde will denote Fourier transform). 
A (comoving) scale of nonlinearity, $r_{\rm nl} (t)$, is defined by the
condition
\begin{displaymath}
  \int_{k < 2 \pi/r_{\rm nl}} \!\! d{\bf k} \; P(k, t) = 1 .
\end{displaymath}

In the self--similar model, the physical properties which do not depend
on the short and large distance cutoffs of $P(k)$
exhibit a simple scaling behavior \citep{Peeb80,Padm95}. The only physically
relevant parameters on which they can depend are $\varrho_b (t_0)$
(through the cosmological background), $a r_{\rm nl} (t_0)$ (through the
initial conditions), and the gravitational constant $G$ (through the
dynamics); self--similarity means that the time $t_0$ is arbitrary.
Thus, in combination with dimensional analysis\footnote{Here it is to
  be distinguished between the dimensions of comoving and physical
  lengths, differing by a factor $a(t)$.}, it is shown that $\sigma(L,
t)$ is a function of the single quantity $L/r_{\rm nl}(t)$; in the linear
regime, $L \gg r_{\rm nl}$, one has $\sigma^2 \propto (L/r_{\rm nl})^{-n-3}$.
Similarly, ${\kappa}_{\rm int/mac}$ can be written in the following
suggestive form:
\begin{equation}
  \label{eq:scaling}
  {\kappa}_{\rm int/mac} (\varrho, t; L) = \varrho_b (\dot{a} r_{\rm nl})^2 
  F_{\rm int/mac} \left( \delta, \frac{L}{r_{\rm nl}} \right) .
\end{equation}
Here $F(\cdot)$ is a dimensionless function of its dimensionless
arguments, and $\varrho_b (\dot{a} r_{\rm nl})^2$ is of the order of the
Hubble--flow kinetic energy in balls of (comoving) radius $r_{\rm nl}$.
The unconditioned averages $\langle \ki \rangle = \langle K_{\rm int}
\rangle$ and $\langle K_{\rm mac} \rangle$ follow the same scaling but
without the $\delta$--dependence.  Deviations from this scaling
behavior would mean a dependence on extra variables, e.g., on short or
large distance cutoffs in $P(k)$ or in the dynamics (as occurs with,
e.g., numerical simulations).

The task now is to characterize the functions $F_{\rm int/mac} (\cdot)$. 
Let $\lambda$ and $R$ denote the short and the
large distance cutoffs, respectively,
so that we take $\lambda \ll r_{\rm nl}, L \ll R$. Roughly speaking, we can
say that $\km$ is determined by the motion at scales between $L$ and
$R$, whereas $\ki$ is dominated by the motion at scales between
$\lambda$ and $L$. 

The hierarchical, bottom--up scenario exhibits a monotonically growing
length scale,
$r_{\rm nl}(t)$, which 
is roughly proportional to the size of the largest collapsed clusters at time
$t$. 
The bottom--up growth of structure by self--gravity can be sketched in
the following picture:
particles get trapped in clusters so that (i) the evolution above the
cluster scale is dominantly ruled by the motion of each cluster as a
whole ( = ``effective particles'') in the gravitational field of the
other clusters, and (ii) the evolution below the cluster scale is
driven mainly by the scales $\sim$ cluster size. This means a
dynamical decoupling between scales above and below the cluster size,
$\sim r_{\rm nl}$ (this idea has been explored by \citet{Domi00,Domi02} in
order to improve the models of structure formation), and implies that
the short--distance cutoff $\lambda$ is irrelevant. Depending on which
of the two motions, (i) or (ii), contributes mostly to the particle
velocity, there arise several possibilities (see Fig.~\ref{fig:sketch}):
\begin{figure}
  \resizebox{\hsize}{!}{\includegraphics{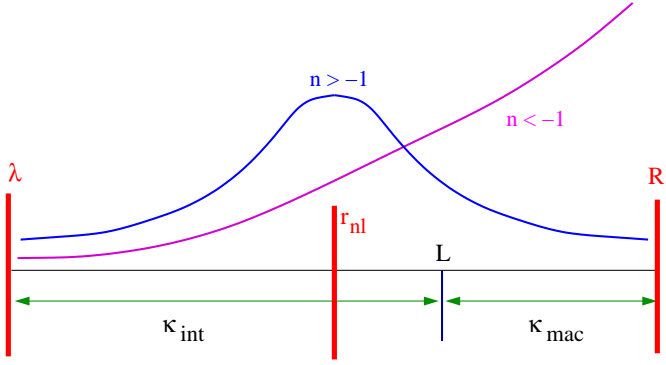}}
  \caption{
    Sketch representing the relative contribution of the different
    length scales to $\km$ and $\ki$. The scales $\lambda$, $R$ and
    $r_{\rm nl}$ are intrinsic to the system, the movable length $L$
    is the observation resolution.}
  \label{fig:sketch}
\end{figure}
\begin{itemize}
\item If $-3<n<-1$, it will be argued that the largest scales dominate.
  Thus, $\km$ should be determined by the scales $\sim R$ and $\ki$ by
  the scales $\sim L$.
\item If $-1<n<4$, the scales around $r_{\rm nl}$ provide the prevalent
  contribution because $\lambda$ is irrelevant and there is no other
  privileged scale between $\lambda$ and $r_{\rm nl}$. We can distinguish
  in turn two cases:
  \begin{itemize}
  \item Linear regime, $r_{\rm nl} \ll L$: the value of $\km$ is
    mainly set by scales $\sim L$, and that of $\ki$ by scales $\sim
    r_{\rm nl}$.
  \item Nonlinear regime, $L \ll r_{\rm nl}$: now $\km$ is dominated by
    scales $\sim r_{\rm nl}$, and $\ki$ by scales $\sim L$.
    \end{itemize}
\end{itemize}
This description will be now elaborated in somewhat more detail. \\

\noindent
{\it Case I: $-3<n<-1$}. There is so much power initially at the large
scales, that the contribution of the linear modes ($r_{\rm nl} k\ll 1$) to
the variance of the macroscopic velocity diverges if $n<-1$:
\begin{displaymath}
  \langle |{\bf u}|^2 \rangle \propto \int_0^{r_{\rm nl}^{-1}} d{\bf k}\ k^{n-2} + 
  \textrm{nonlinear contribution} .
\end{displaymath}
Hence, both ${\bf u}$ and the particle velocities ${\bf u}_\alpha$
will be mainly determined by the modes around the largest available
scale, the infrared (IR) cutoff $R$.
For the macroscopic kinetic energy we then estimate
\begin{equation}
  \label{eq:mac2}
  \km \sim \varrho \langle |{\bf u}|^2 \rangle_\varrho \sim
  \varrho_b (\dot{a} R)^2 \sigma^2 (R) (1+\delta) , 
\end{equation}
\begin{equation}
  \label{eq:meanmac2}
  \langle K_{\rm mac} \rangle \sim \varrho_b (\dot{a} R)^2 \sigma^2 (R) .
\end{equation}
For the internal kinetic energy, it proves useful to consider
separately the linear and nonlinear regimes:
\begin{itemize}
\item Linear regime, $L \gg r_{\rm nl}$. Our hypothesis is that the main
  contribution to $\ki$ comes from scales $\sim L$, so that we can
  employ the linear solutions to the gravitational instability to
  evaluate Eqs.~(\ref{eq:fields}). The calculations are collected
  in the Appendix; the final result reads
  \begin{equation}
    \label{eq:intlin2}
    \ki = \langle K_{\rm int} \rangle \left[ 1 + B \left( 
        1 - \frac{\delta^2}{\sigma^2(L)} \right) \right] .
  \end{equation}
  In this expression, $\langle K_{\rm int} \rangle$ is the unconditioned
  average internal peculiar kinetic energy in the linear
  approximation, Eq.~(\ref{eq:meankin}), $B$ is a dimensionless
  negative numerical coefficient, Eq.~(\ref{eq:Bcoeff}). Both $\langle
  K_{\rm int} \rangle$ and $B$ are IR--convergent because the
  $R$--dependence cancels from the difference ${\bf u}_\alpha-{\bf u}$,
  so that Eq.~(\ref{eq:scaling}) holds. Indeed, we derive the scaling
  behavior
  \begin{equation}
    \label{eq:meanintlin2}
    \langle K_{\rm int} \rangle \propto \varrho_b (\dot{a} L)^2 \sigma^2 (L).
  \end{equation}
  
\item Nonlinear regime, $L \ll r_{\rm nl}$. In high density coarsening cells,
  $\delta \gg 1$, the Hubble flow is negligible and peculiar
  velocities are approximately equal to the physical velocities.
  Assuming stationarity, the conditioned differences $\langle |{\bf
    u}_\alpha - {\bf u}|^2 \rangle_\varrho$ can be expected to be given
  by local virialization on scales $\sim L$, so that
  \begin{equation}
    \label{eq:virial}
    \ki \sim \frac{1}{(a L)^3} \frac{G [\varrho (a L)^3]^2}{a L} 
    \sim \varrho_b (\dot{a} L)^2 \delta^2, \quad (\delta \gg 1) .
  \end{equation}
  The opposite limit of low density, $0<1+\delta \ll 1$, requires a
  model for the expansion of voids. The simplest model would set
  $|{\bf u}_\alpha - {\bf u}| \sim h L$ typically, with $h \approx
  \dot{a}$ quantifying the void expansion speed. Then
  \begin{equation}
    \label{eq:lowdens}
    \ki \sim \varrho_b (\dot{a} L)^2 (1+\delta) , 
    \qquad (0< 1+\delta \ll 1) ,
  \end{equation}
  predicting a low velocity dispersion. 
  
  To compute the unconditioned average $\langle K_{\rm int} \rangle$, it
  must be noticed that, although the internal kinetic energy of
  high--density cells is very large, the number of low--density cells is
  much larger, and it is not clear which of the two competing effects
  dominates. If one would assume that the main contribution comes from
  high--density cells, then
  Eq.~(\ref{eq:virial}) would yield
  \begin{equation}
    \label{eq:meanvirial}
    \langle K_{\rm int} \rangle \sim \varrho_b (\dot{a} L)^2 \sigma^2 (L) .
  \end{equation}

\end{itemize}
\bigskip

\noindent
{\it Case II: $-1<n<4$}. The linear modes do not lead to divergences
and the velocities are now determined mainly by the scale
$r_{\rm nl}$.
\begin{itemize}
  
\item Linear regime, $L \gg r_{\rm nl}$. Now the linear solution states
  that the smoothed velocity ${\bf u}$ will be determined by scales
  $\sim L$, i.e., the smallest scales it probes, and thus one obtains
 \begin{equation}
   \label{eq:maclin01}
   \km \sim \varrho_b \langle |{\bf u}|^2 \rangle \sim 
   \varrho_b (\dot{a} L)^2 \sigma^2 (L) , \quad 
   \langle K_{\rm mac} \rangle \sim \km .
 \end{equation}
 On the other hand, our hypothesis is that the particle velocities
 ${\bf u}_\alpha$ are controlled by scales $\sim r_{\rm nl}$. Hence, we
 write the estimate
 \begin{displaymath}
   \langle |{\bf u}_\alpha|^2 \rangle_\varrho \sim 
   \frac{G \varrho_b (a r_{\rm nl})^3}{a r_{\rm nl}} \sim (\dot{a} r_{\rm nl})^2 ,
 \end{displaymath}
 assuming that the scales $\sim r_{\rm nl}$ are rather insensitive to the
 small density fluctuations at the scale $L \gg r_{\rm nl}$. Then 
 \begin{equation}
   \label{eq:intlin01}
   \ki \sim \varrho_b (\dot{a} r_{\rm nl})^2 (1 + \delta) , 
   \quad \langle K_{\rm int} \rangle \sim \varrho_b (\dot{a} r_{\rm nl})^2.
 \end{equation}
 
\item Nonlinear regime, $L \ll r_{\rm nl}$. The conditioned difference
  $\langle |{\bf u}_\alpha - {\bf u}|^2 \rangle_\varrho$ is controlled
  by the local dynamics on scales $\sim L$ like in the previous case,
  and expressions~(\ref{eq:virial}--\ref{eq:meanvirial}) are valid now
  too.  The motion of the coarsening cells as a whole, and thus
  $\langle |{\bf u}|^2 \rangle_\varrho$, is however assumed to be
  determined by the large scales $\sim r_{\rm nl}$, so that
  \begin{equation}
    \label{eq:macnonlin01}
    \km \sim \varrho_b (\dot{a} r_{\rm nl})^2 (1+\delta) , \quad
    \langle K_{\rm mac} \rangle \sim \varrho_b (\dot{a} r_{\rm nl})^2 ,
  \end{equation}
  in analogy to Eqs.~(\ref{eq:intlin01}).
\end{itemize}

\section{Simulation and analysis method} 
\label{sec:method}

The simulations of the self--similar models are described in full
detail elsewhere \citep{MeSh93}. They consist of a cubic box (periodic
boundary conditions) of comoving sidelength $R$ containing
$N=128^3$ particles.
The dynamical evolution was computed using a PM algorithm on a grid
with Nyquist wavenumber $=128 \pi/R$. The background cosmological
expansion followed the Einstein--de Sitter solution and the initial
conditions were generated by using the Zel'dovich approximation for a
Gaussian random field with a scale--invariant power spectrum $P(k)
\propto k^n$.  In the present work, the values $n=-2, 0, +1$ were considered
and the data were studied at three different times,
corresponding to a scale of nonlinearity
$r_{\rm nl} = R/64$, $r_{\rm nl} = R/16$, and $r_{\rm nl} = R/4$, respectively. In each case
four independent realizations of the initial conditions were evolved.

Two CDM models were also addressed with a PM algorithm: flat CDM
($\Omega_m = 1.0$, $\Omega_\Lambda = 0.0$, $\sigma_8 \approx 1$), and
open CDM ($\Omega_m = 0.3$, $\Omega_\Lambda = 0.0$, $\sigma_8 \approx
1$). Each simulation contained $N=256^3$, and for each model two
different box--sizes were considered, $R=128$~Mpc and $R=512$~Mpc.

Starting from the coordinates $\{ {\bf x}_\alpha, {\bf u}_\alpha
\}_{\alpha=1 \dots N}$ provided by the simulation, the definitions
(\ref{eq:fields}) were applied with a cubic top--hat window,
\begin{equation}
  \label{eq:window}
  W({\bf z}) = \theta (1 - 2 |z_1|) \; \theta (1 - 2 |z_2|) \; 
  \theta (1 - 2 |z_3|) ,
\end{equation}
where $\theta(\cdot)$ is the step function. The coarsening was
implemented efficiently by covering the simulation box with a cubic
grid; in order to probe a minimum of $2048$ coarsening cells, the grid
was shifted randomly, when needed. It was checked that the results are
insensitive to the use of spherical top--hat windows instead. A
comparison with the use of a Gaussian window was also carried out. We
observed a difference only for the case $n=+1$ at the earliest time,
$r_{\rm nl} = R/64$, because then $\sigma^2$ diverges for a top--hat window
(in the simulation, the artificial point--particle discreteness regularizes the
singularity). Hence, for this single case we disregarded the use of the cubic
top--hat window as unphysical and used instead a Gaussian window (which
is computationally much more costly).

The explored values of the coarsening length $L$ were equally
separated in a logarithmic scale and they ranged from a maximum
$(R/5)$ down to a minimum $\approx \ell/2$, where $\ell := R/N^{1/3}$
is the average interparticle distance.
This results in scatter plots ``kinetic energy vs. density''. To
compute the constrained average~(\ref{eq:eqstate}), the data for the
kinetic energy
were binned into 40 subintervals according to the value of the density
$\varrho$; bins containing less than 10 data points were disregarded.
It was checked that the conclusions are robust against the amount of
binning by varying the number of bins.
The constrained average was identified with the mean of each bin.
The amount of scatter about this (global) mean is represented in the
plots by scatter bars, which extend between the mean of those
kinetic energies {\em above} the global mean, and the mean of
those kinetic energies {\em below} the global mean. We find that
this method represents the scatter of the data in the log--plots more
faithfully than the estimation through the variance.

We checked the algorithm in various ways. It was applied to an ideal
gas simulation: the results for the dependence of $\ki$ on
the density agreed with the ideal gas equation of state. Another check
was to restrict the coarsening procedure to a subvolume of the
simulation box (1/64 of the total volume) for some sample cases.
As expected, we find that the data are somewhat noisier
because of the reduced number of particles, but the conclusions remain
the same.

\section{Results} 
\label{sec:result}

There are a couple of general remarks which hold for all the Subsecs.~to follow.
First, the small length scale $\ell$ enters in the results via
mass--resolution and force--resolution effects: the first effect refers to
the presence of a minimum non--vanishing mass --- that of a single
particle.
This affects the computation of the unconstrained averages $\langle
K_{\rm int/mac} \rangle$ due to undersampling of the cells with a mass
smaller than this minimum, which also sets a lower bound on the value
of the density at a given fixed $L$ when computing the constrained
averages, $\ki$, $\km$. In particular, it renders all results
concerning the nonlinear regime ($L<r_{nl}$) at the earliest probed
time ($r_{\rm nl}=R/64$) rather unreliable, since then $r_{nl}
\approx \ell$. The second effect, force resolution, implies that the
relative force over two particles decreases when they are closer than
the mesh spacing of the PM algorithm, $\approx \ell$, and the velocity
dispersion below this scale does not grow as much as it would if $\ell
\rightarrow 0$. All in all, these two effects tend to artificially
reduce the value of the kinetic energy, in particular of $K_{\rm
  int}$, being more sensitive to the small scales. The theoretical
discussion in Sec.~\ref{sec:theory} suggests this effect to be
particularly noticeable when $n>-1$ and at the earliest times, as will
be indeed observed.

Second, the influence of the ``cosmic variance'', i.e., of the
fluctuations in measured quantities from one realization to another,
is the strongest when $n=-2$. For clarity, however, we will show in
the plots the results of a single realization, since the other ones
yield 
almost identical results.

\subsection{Self--similar models: mass density variance}

\begin{figure}[t]
  \resizebox{\hsize}{!}{\includegraphics{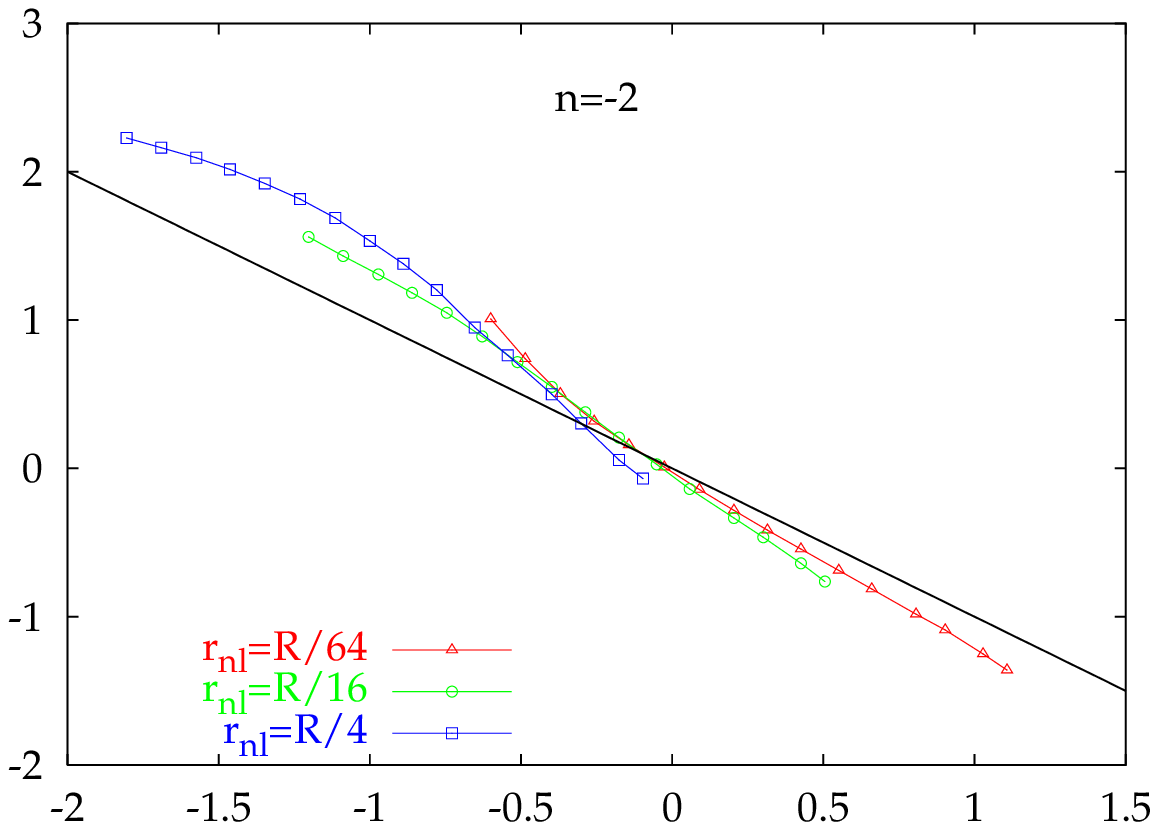}}
  \resizebox{\hsize}{!}{\includegraphics{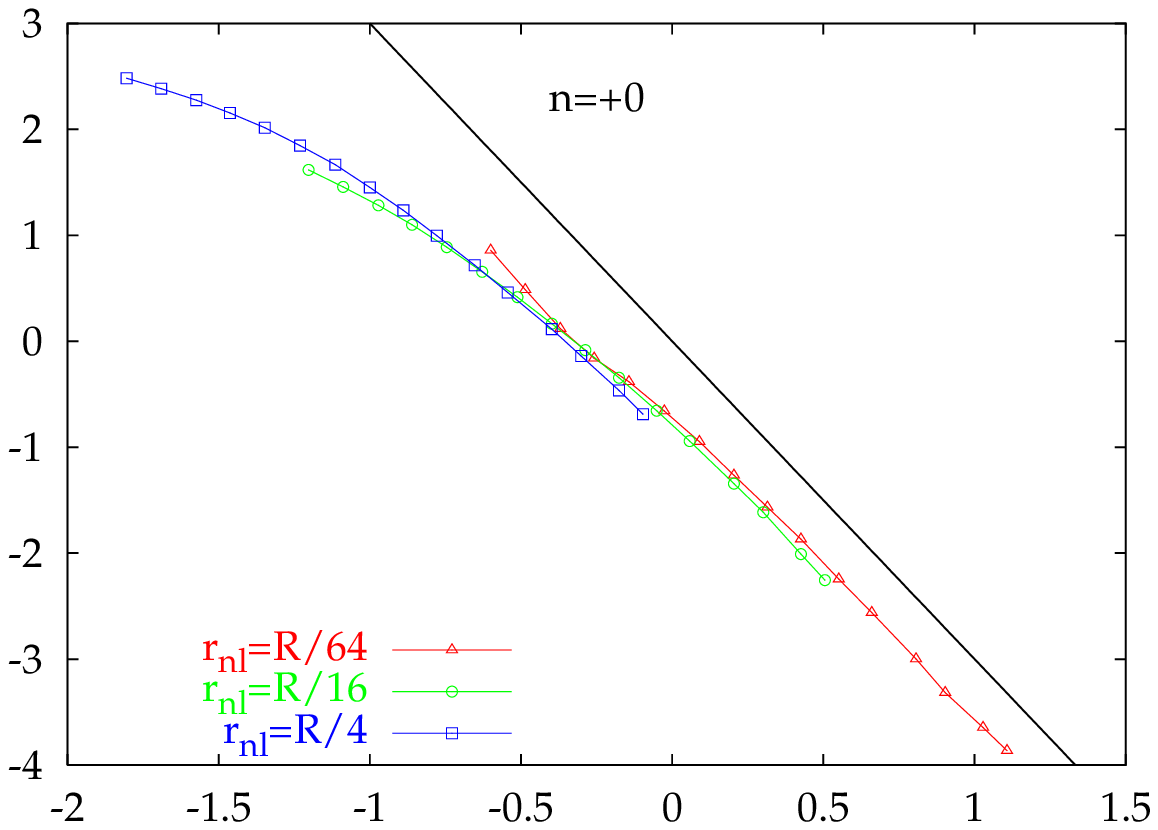}}
  \resizebox{\hsize}{!}{\includegraphics{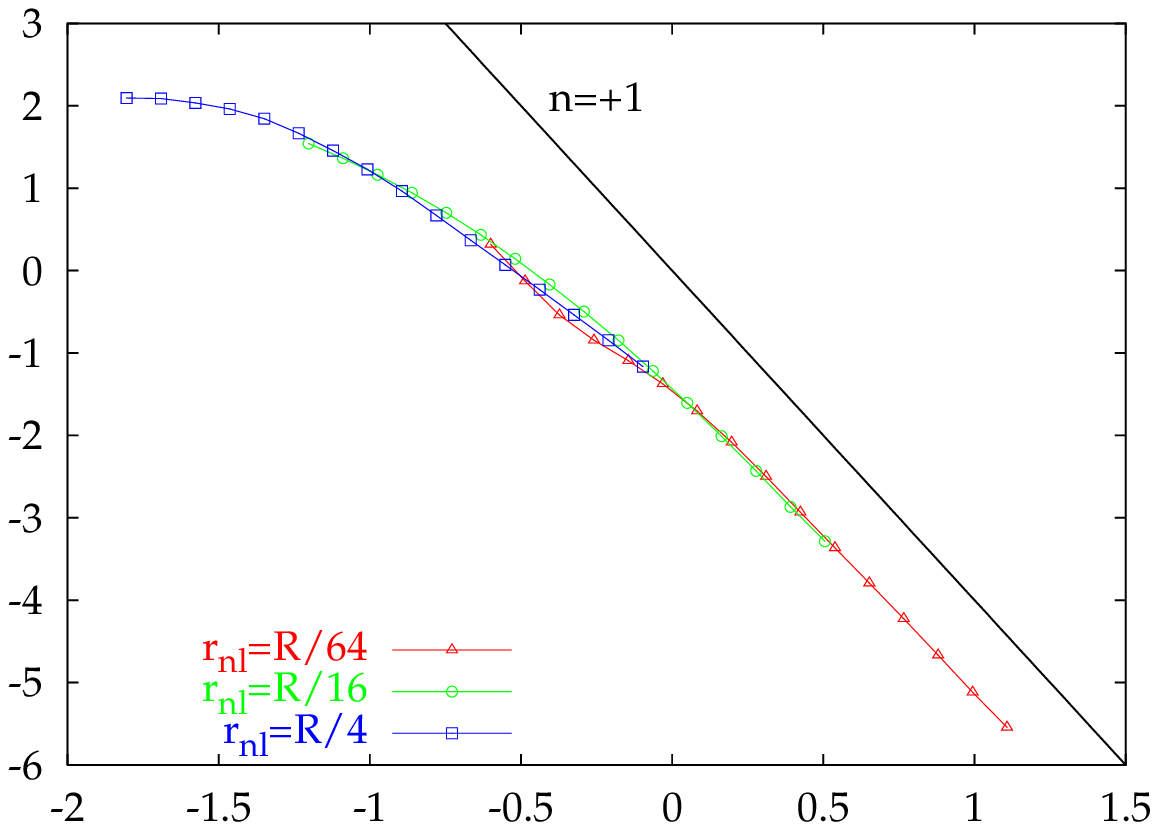}}
  \caption{
    Abscissa: $\log(L/r_{\rm nl})$. Ordinate: $\log \sigma^2 (L)$ at the
    three probed times. Solid line: expected functional dependence in
    the linear regime, $\sigma^2 \propto L^{-n-3}$.  }
  \label{fig:sigma}
\end{figure}

For reference purposes, Fig.~\ref{fig:sigma} shows the measurements of
$\sigma^2(L)$. The results collapse well on a single function of
$L/r_{\rm nl}$. At the earliest time and the smallest lengths, one can
observe the beginning of the crossover to the Poissonian behavior,
$\sigma^2(L) \propto L^{-3}$, induced by the small--scale discreteness.
At large $L$, one recovers the linear scaling, $\sigma^2(L) \propto
L^{-n-3}$; due to finite--size effects, the case $n=-2$ exhibits a
slight departure away from this dependence.

\subsection{Self--similar models: the unconstrained averages}

\begin{figure}[t]
  \resizebox{\hsize}{!}{\includegraphics{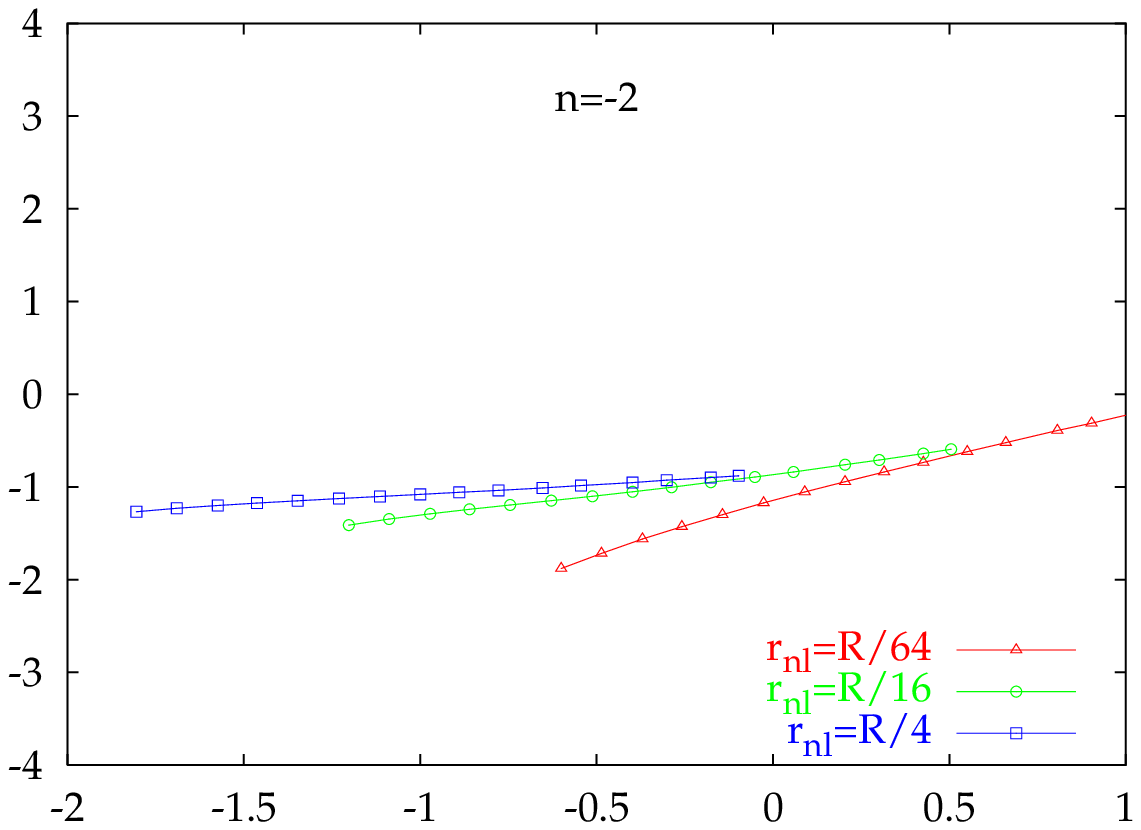}}
  \resizebox{\hsize}{!}{\includegraphics{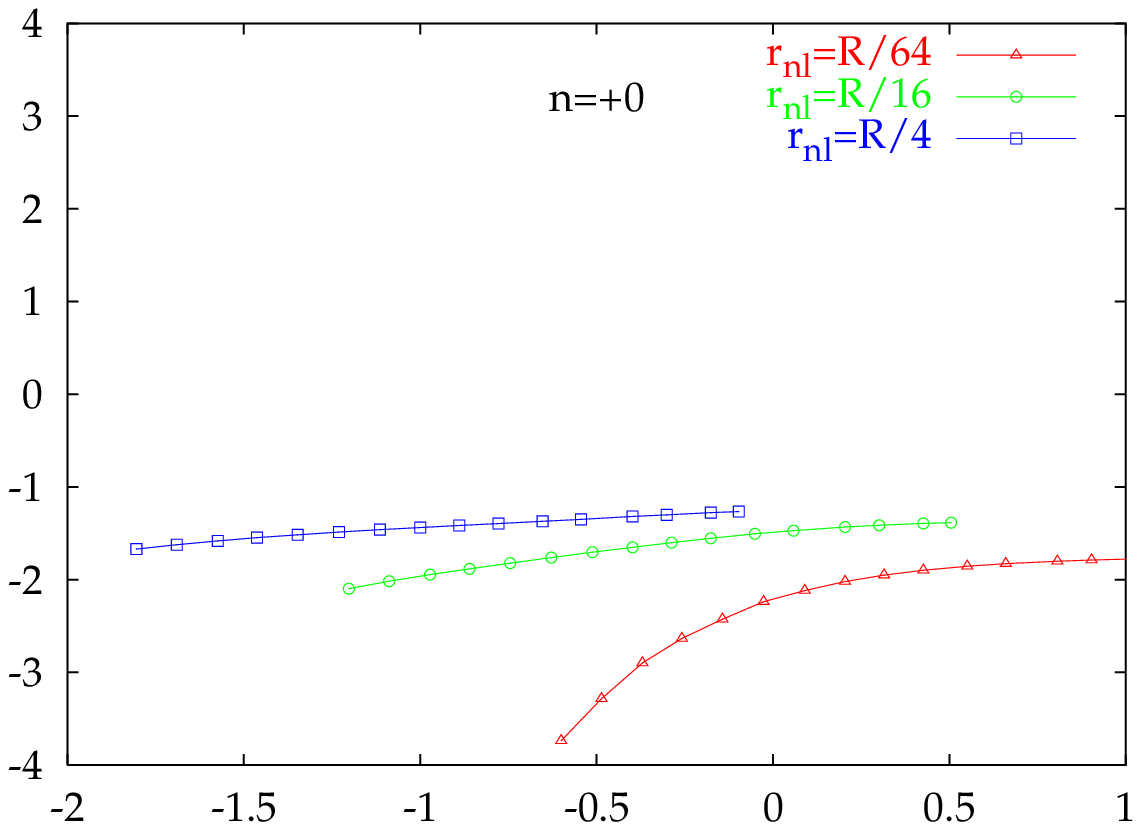}}
  \resizebox{\hsize}{!}{\includegraphics{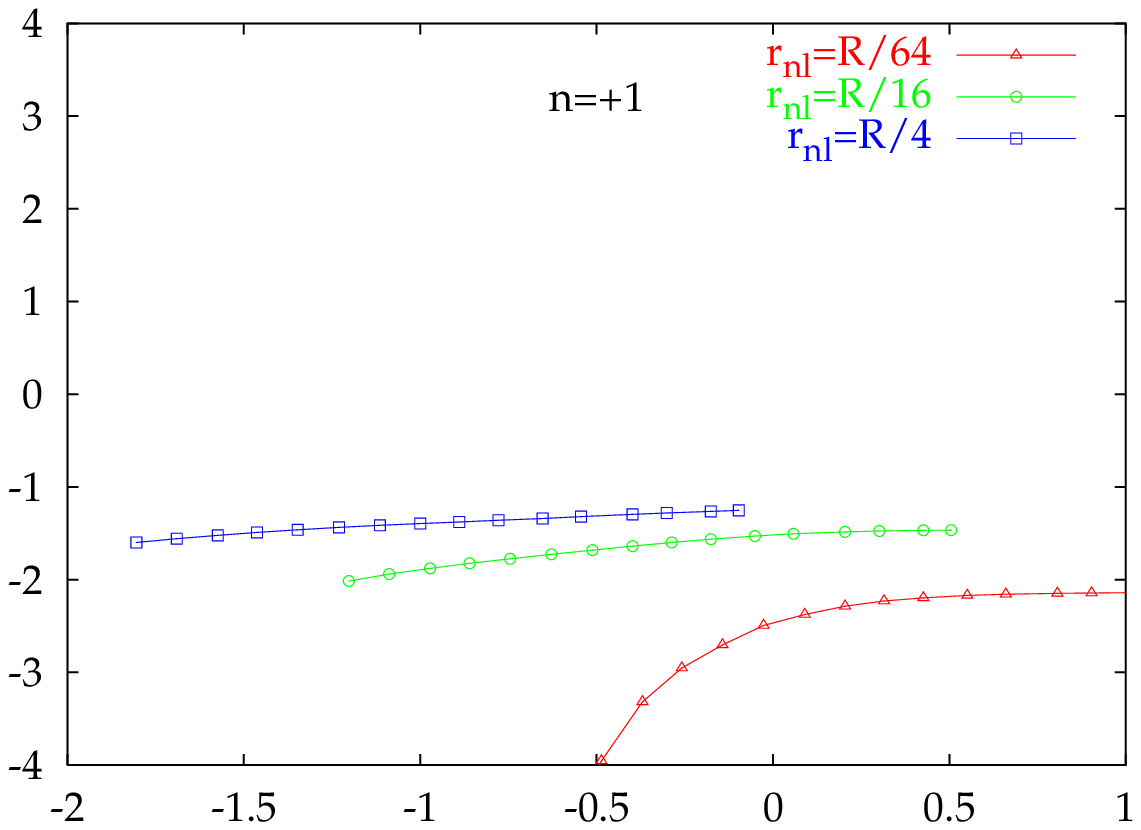}}
  \caption{
    Abscissa: $\log(L/r_{\rm nl})$. Ordinate: $\log[\langle K \rangle
    / \varrho_b (\dot{a} r_{\rm nl})^2]$ at the three probed times:
    the lower points in each plot correspond to $\langle K_{\rm int}
    \rangle$; the upper points represent $10^3 \langle K_{\rm mac}
    \rangle$. The solid line is proportional to $L^2 \sigma^2 (L)$,
    Eqs.~(\ref{eq:meanintlin2},\ref{eq:meanvirial},\ref{eq:maclin01}).}
  \label{fig:meankin}
\end{figure}

The average $\langle K_{\rm mac} \rangle$ is the kinetic equivalent of
$\sigma^2$. In Fig.~\ref{fig:meankin}, we observe that the data for
$\langle K_{\rm mac} \rangle$ do not follow at all the scaling
behavior~(\ref{eq:scaling}) when $n=-2$. As explained in
Sec.~\ref{sec:theory}, this is due to finite--size effects: we have
checked that the data follow instead the dependence~(\ref{eq:meanmac2}).
The data for the other two cases, $n=0,+1$ on the contrary, obey the
expected scaling, Eq.~(\ref{eq:maclin01}) when $L \gg r_{\rm nl}$, and
Eq.~(\ref{eq:macnonlin01}) when $L \ll r_{\rm nl}$. A major departure in
the three cases is observed at the earliest time ($r_{\rm nl}=R/64$) in the
nonlinear regime ($L < r_{\rm nl}$) 
due to the undersampling problem mentioned above.
This was confirmed by artificially removing from the estimate of the
averages those cells with less than a given number of particles, yielding
the same behavior in $\langle K_{\rm mac} \rangle$ as detected in the
plots.

As remarked in Sec.~\ref{sec:theory}, the average $\langle K_{\rm int}
\rangle$ is more sensitive to the small--scale dynamics than $\langle
K_{\rm mac} \rangle$ or $\sigma^2$ are. This average also suffers
the same undersampling problem as $\langle K_{\rm mac} \rangle$. But
the resolution effects are somewhat larger and prevent the data from
following the scaling~(\ref{eq:scaling}) perfectly. One can recognize
a tendency for these effects to become less relevant in time, and to
be more important for larger values of the spectral index $n$, in
agreement with the theoretical discussion.

\subsection{Self--similar models: the constrained averages}

\subsubsection{Check for self--similarity}

\begin{figure}[t]
  \resizebox{\hsize}{!}{\includegraphics{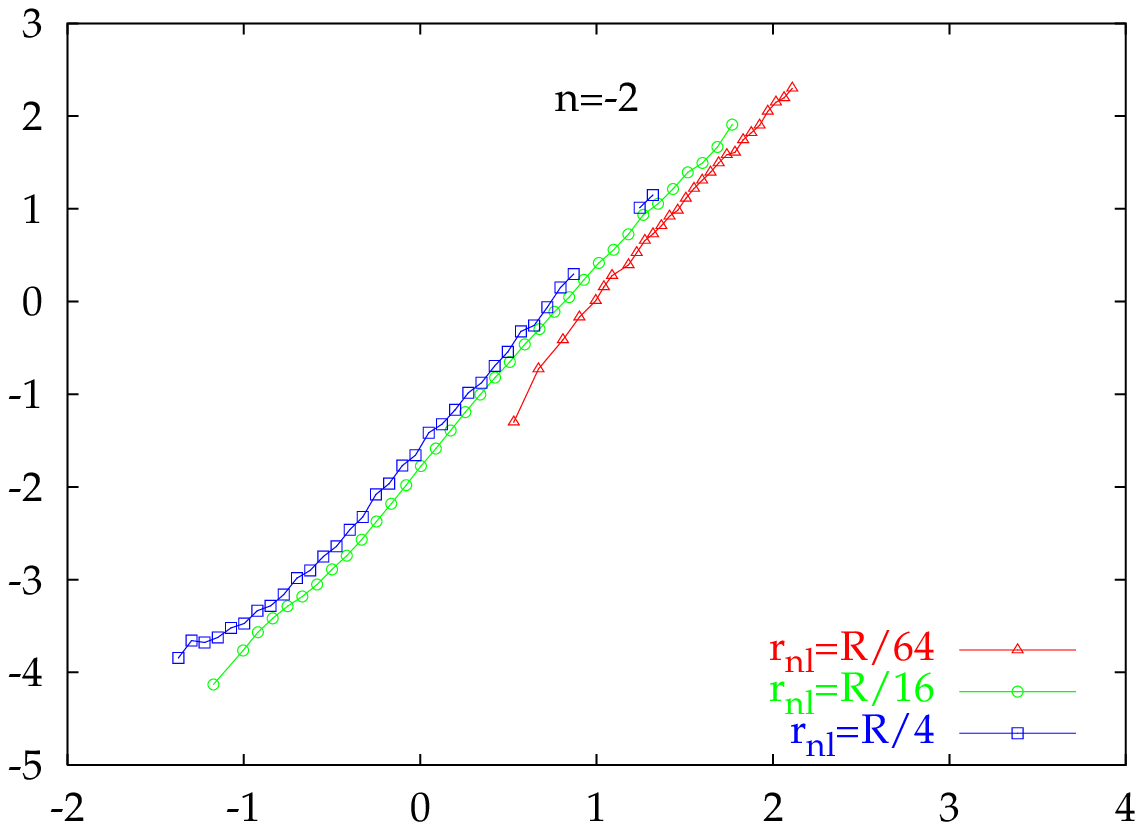}}
  \resizebox{\hsize}{!}{\includegraphics{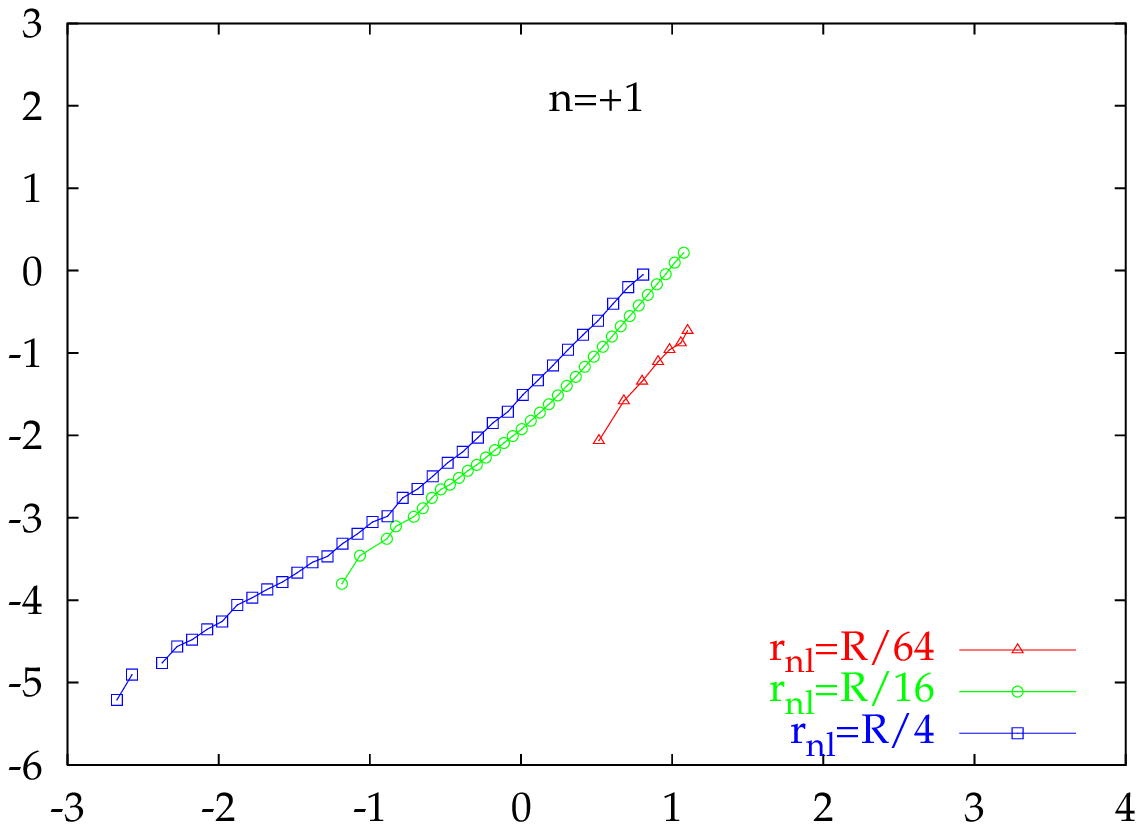}}
  \caption{
    Abscissa: $\log(1+\delta)$. Ordinate: $\log[\ki/\varrho_b (\dot{a}
    r_{\rm nl})^2]$. The plots are arbitrarily shifted in horizontal
    direction for clarity. From left to right, $L/r_{\rm nl} \approx
    2.67, 0.40, 0.08$. When self--similarity holds,
    Eq.~(\ref{eq:scaling}), curves corresponding to the same ratio
    $L/r_{\rm nl}$ fall on top of each other.}
  \label{fig:scaling}
\end{figure}

We first checked if the measured $\kappa_{\rm int/mac}$ followed the
self--similar scaling relationship~(\ref{eq:scaling}). The conclusions
are almost the same as derived above with the unconstrained averages
$\langle K_{\rm int/mac} \rangle$: $\km$ follows self--similarity very
well (except if $n=-2$), while $\ki$ follows it a bit less well. The
important difference is that departures from self--similarity are
(sometimes substantially) smaller than in $\langle K_{\rm int/mac}
\rangle$, see Fig.~\ref{fig:scaling}. The reason is that $\kappa_{\rm
  int/mac}$ suffer the undersampling problem due to finite
mass--resolution only in the small--$\delta$ end of each curve or at the earliest time. When the
number of particles in the cell is large enough, force--resolution is
likely the main effect and it does not appear to spoil self--similarity so much.

\subsubsection{$\km$}
  
\begin{figure}[t]
  \resizebox{\hsize}{!}{\includegraphics{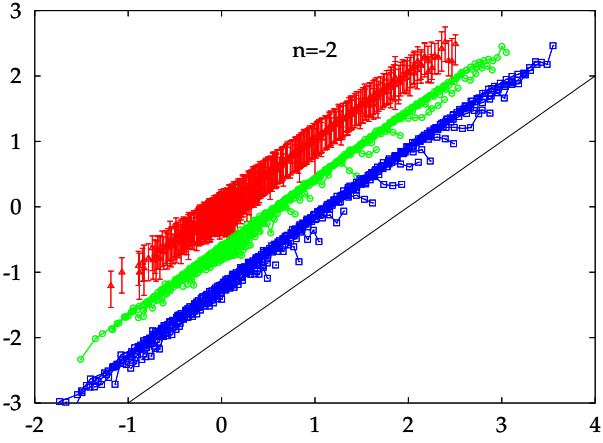}}
  \caption{
    Abscissa: $\log(1+\delta)$. Ordinate: $\log[\km / \varrho_b
    (\dot{a} r_{\rm nl})^2]$. Solid line: theoretical functional
    dependence, Eq.~(\ref{eq:mac2}). All the probed coarsening lengths, 
    at the three probed times (increasing from top to bottom) are
    plotted. The scatter bars are barely dependent on time, coarsening
    length and density; for clarity, they are shown only at the
    earliest time.
    }
  \label{fig:mac2}
\end{figure}

Fig.~\ref{fig:mac2} shows the function $\km (\delta)$ at the different
times and coarsening lengths probed for the spectral index $n=-2$. The
scaling Eq.~(\ref{eq:mac2}) is obeyed well at all times, although we
find small fluctuations around the linear
dependence between realizations. This is due to
the strong dependence on the IR cutoff. It is also responsible for the
lack of collapse of the three plots on a single function, the deviations
following indeed the law $R^2 \sigma^2(R)$ of the factor in
Eq.~(\ref{eq:mac2}). 
The slight deviations around $\delta=0$, most noticeable at the
earliest time, correspond to the largest values of L and are
likely due to finite--size effects too.

\begin{figure}[h]
  \resizebox{\hsize}{!}{\includegraphics{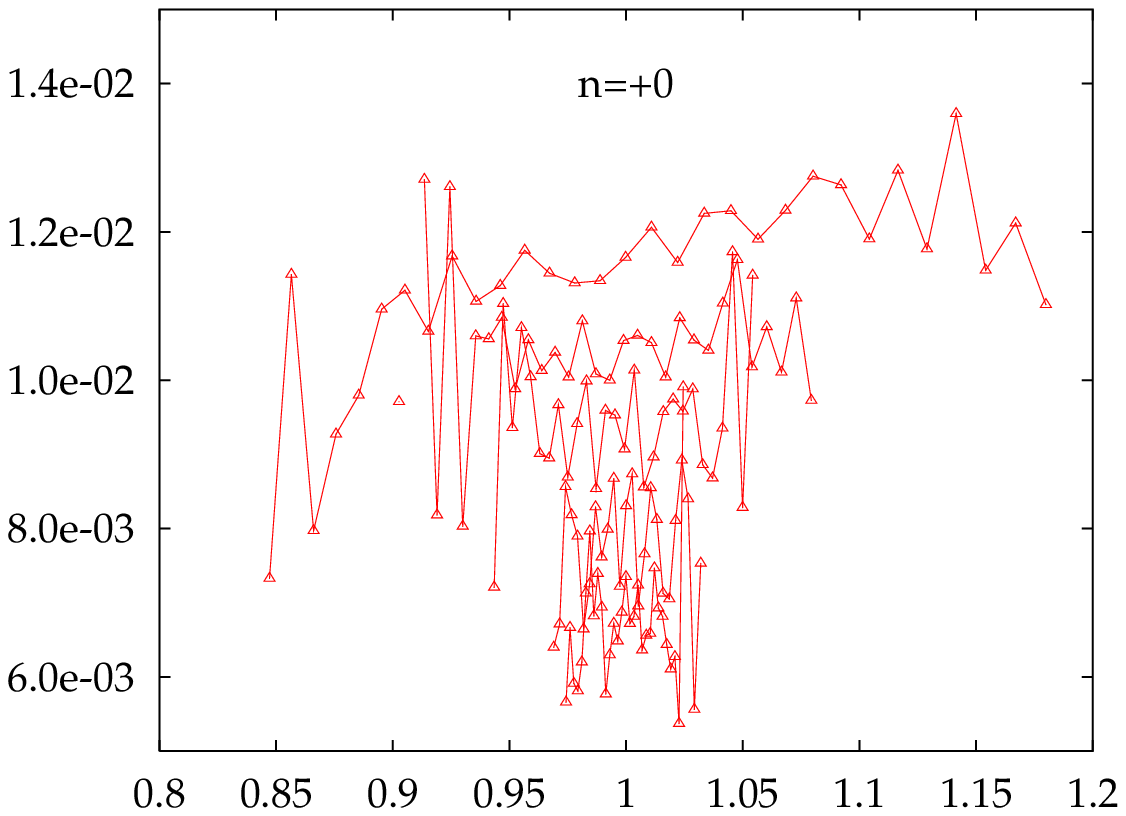}}
  \resizebox{\hsize}{!}{\includegraphics{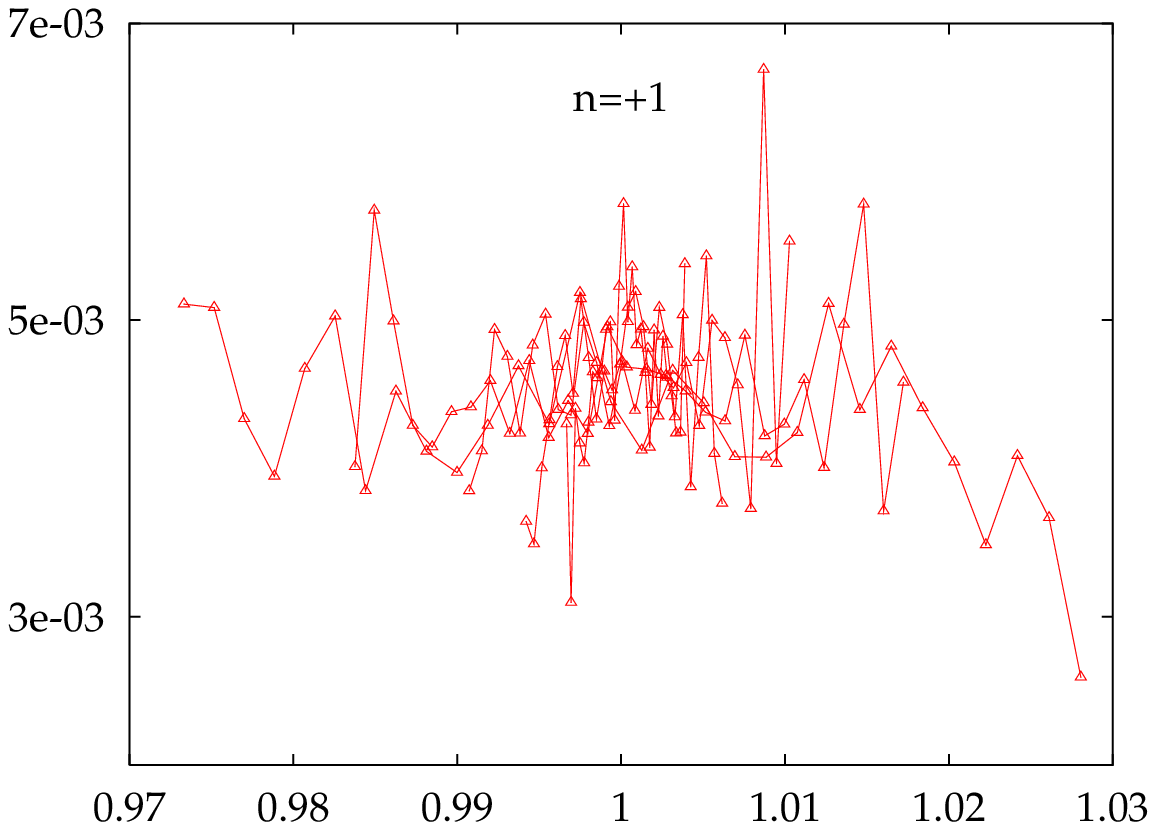}}
  \caption{
    Abscissa: $1+\delta$. Ordinate: $\km / \varrho_b
    [\dot{a} L \sigma(L)]^2$. Only plotted are the coarsening
    lengths such that $\sigma(L)<0.05$ if $n=0$ and $\sigma(L)<0.03$ if
    $n=+1$.
    The theoretical functional dependence, Eq.~(\ref{eq:maclin01}), is a constant.
    The scatter bars extend beyond the plotted area.}
  \label{fig:maclin01}
\end{figure}

\begin{figure}[h]
  \resizebox{\hsize}{!}{\includegraphics{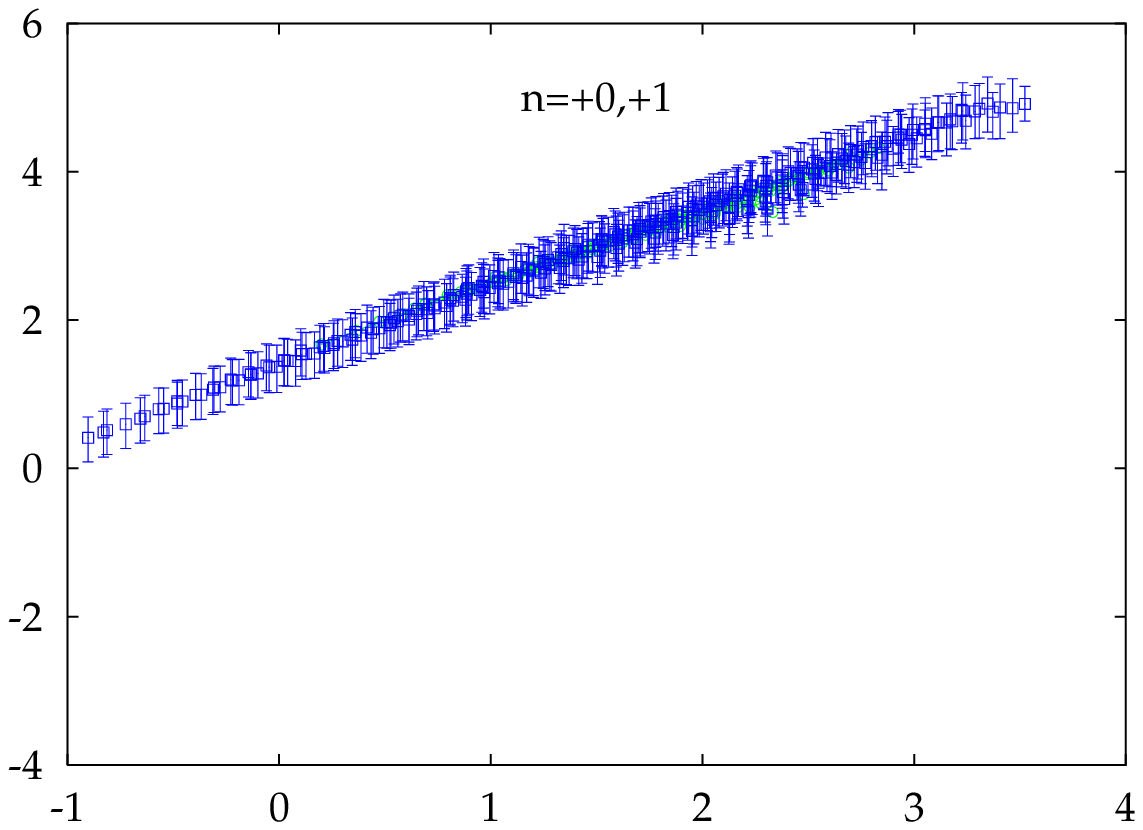}}
  \caption{
    Abscissa: $\log(1+\delta)$. Ordinate: $\log[\km / \varrho_b
    (\dot{a} r_{\rm nl})^2]$. ($\km$ for the case $n=0$ is shifted by a
    factor $10^3$). Solid line: theoretical functional dependence,
    Eq.~(\ref{eq:macnonlin01}). Only plotted are the coarsening
    lengths such that $\sigma(L)>\sqrt{10}$. The scatter bars are barely
    dependent on time, coarsening length and density; for clarity,
    they are shown only for $n=0$.}
  \label{fig:macnonlin01}
\end{figure}

Fig.~\ref{fig:maclin01} shows $\km (\delta)$ in the linear regime when
$n=0,+1$.  The behavior~(\ref{eq:maclin01}) is obeyed very well within
the scatter bars; in the case $n=0$, a systematic trend away from the
expected data collapse is observed.
Fig.~\ref{fig:macnonlin01} corresponds to the nonlinear regime.  The
theoretical scaling Eq.~(\ref{eq:macnonlin01}) is also very well
followed within scatter bars. For the largest values of $L$ considered
in the plot, a slight tendency is noticeable away from the theoretical
dependence, which is more obvious for $n=+1$. 

 
\subsubsection{$\ki$, linear regime}

\begin{figure}
  \resizebox{\hsize}{!}{\includegraphics{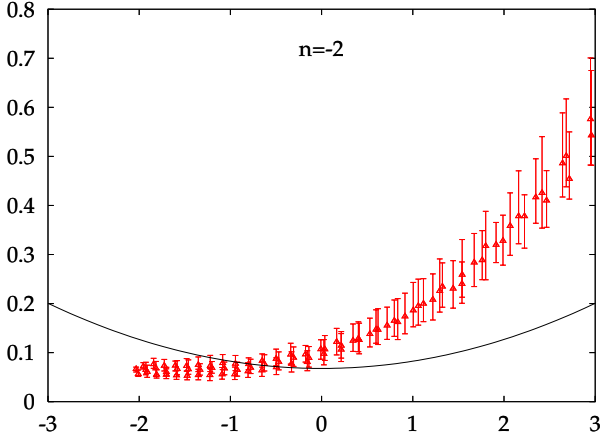}}
  \caption{
    Abscissa: $\delta/\sigma(L)$. Ordinate: $\ki /
    \varrho_b [\dot{a} L \sigma(L)]^2$. Solid line: theoretical
    prediction, Eq.~(\ref{eq:intlin2}). Only coarsening lengths are
    plotted such that $\sigma(L)<0.3$.  }
  \label{fig:intlin2}
\end{figure}

Fig.~\ref{fig:intlin2} shows $\ki (\delta)$ for $n=-2$ in the linear
regime. Deviations from the theoretical prediction,
Eq.~(\ref{eq:intlin2}) with $\langle K_{\rm int} \rangle$ and $B$
computed with Eqs.~(\ref{eq:meankin}, \ref{eq:Bcoeff}),
are noticeable.  The probable reason for
the discrepancy is that the linear regime is not really being probed:
one can observe a large asymmetry $\delta \rightarrow - \delta$ in the
plot, in contradiction with a centered Gaussian distribution for
$\delta$. (That deviations from Gaussianity are indeed important has
been checked by estimating the probability distribution of $\delta$
from the simulations).  The most linear case we probe for $n=-2$
(corresponding to $r_{\rm nl}=R/64$ and $L=R/5$) yields $\sigma
\approx 0.2$, which satisfies the asymptotic condition $\sigma \ll 1$
only marginally.  Thus, the ultimate origin of the problem is the IR
cutoff imposed by the simulation box which limits the maximum value of
$L$.

\begin{figure}
  \resizebox{\hsize}{!}{\includegraphics{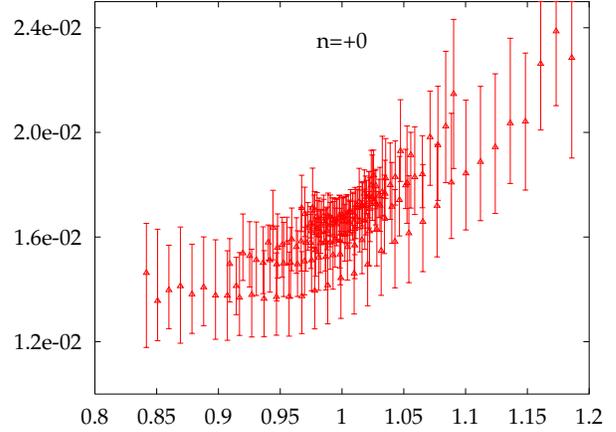}}
  \resizebox{\hsize}{!}{\includegraphics{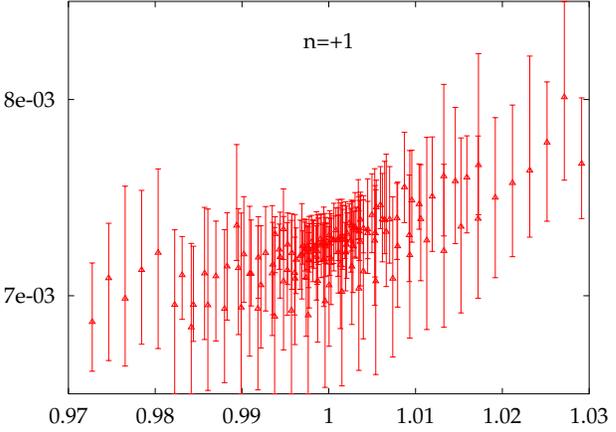}}
  \caption{
    Abscissa: $1+\delta$. Ordinate: $\ki / \varrho_b
    (\dot{a} r_{\rm nl})^2$. Theoretical prediction: a linear dependence.
    Only coarsening lengths such that $\sigma(L)<0.05$ if $n=0$ and
    $\sigma(L)<0.03$ if $n=+1$.
    }
  \label{fig:intlin01}
\end{figure}

Fig.~\ref{fig:intlin01} represents $\ki (\delta)$ when $n=0,+1$.
According to Eq.~(\ref{eq:intlin01}), $\ki$ is determined by the
velocities at scales $r_{\rm nl}$, which is $\sim \ell$ at the
earliest time (corresponding to the plotted data). Thus, the results
are not too reliable in principle, as discussed above. Indeed, the
collapse of the data on a master curve is marginal within the scatter bars,
$\ki$ exhibits a trend to decrease with decreasing coarsening length
 --- consistent with the artificial reduction of kinetic energy at
small scales by resolution effects. The linear dependence predicted by
Eq.~(\ref{eq:intlin01}) is not observed, a curvature being evident.
But we cannot conclude if this is due to resolution effects or because
the derivation of Eq.~(\ref{eq:intlin01}) relies on too simple arguments.

\subsubsection{$\ki$, nonlinear regime}

\begin{figure}
  \resizebox{\hsize}{!}{\includegraphics{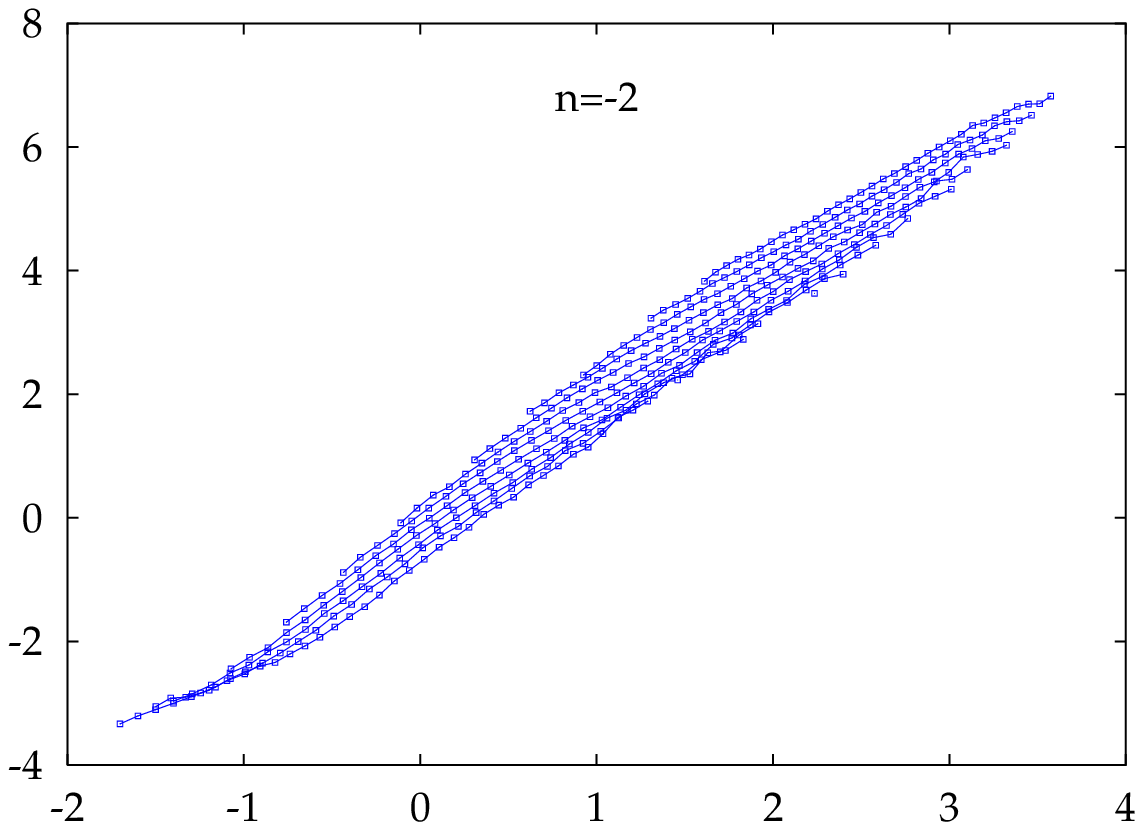}}
  \resizebox{\hsize}{!}{\includegraphics{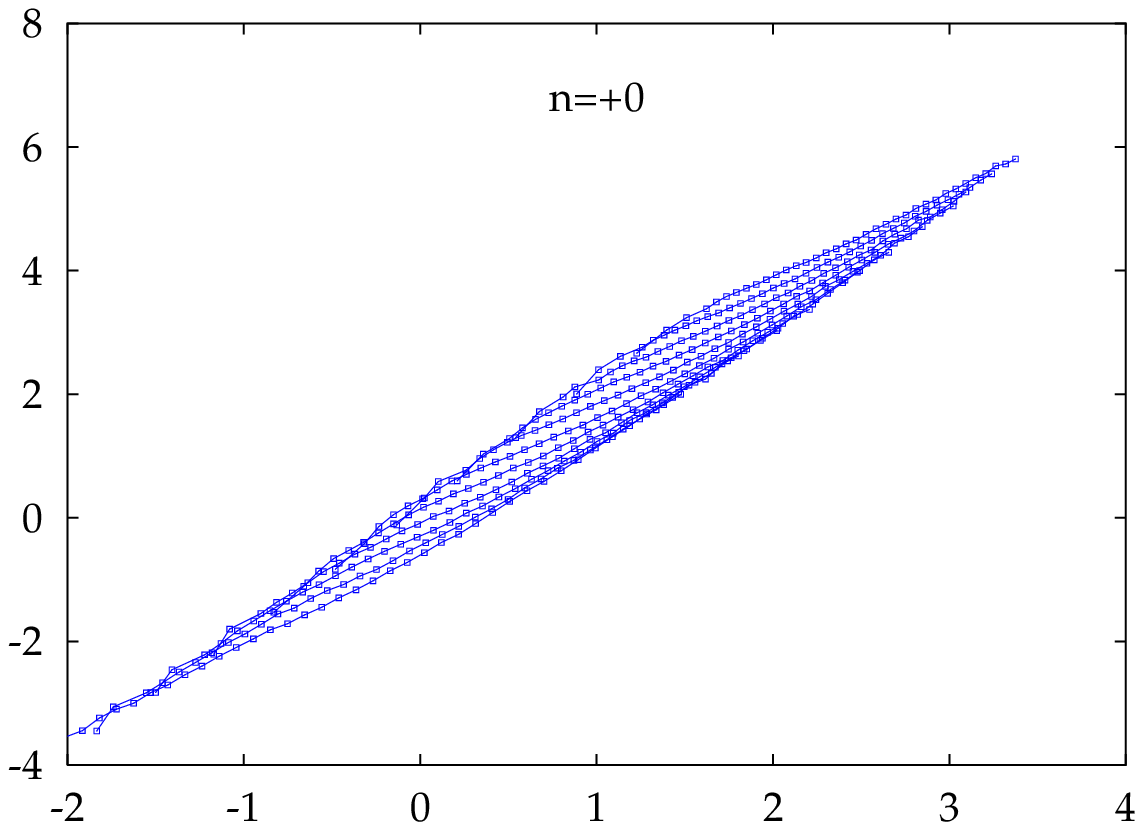}}
  \resizebox{\hsize}{!}{\includegraphics{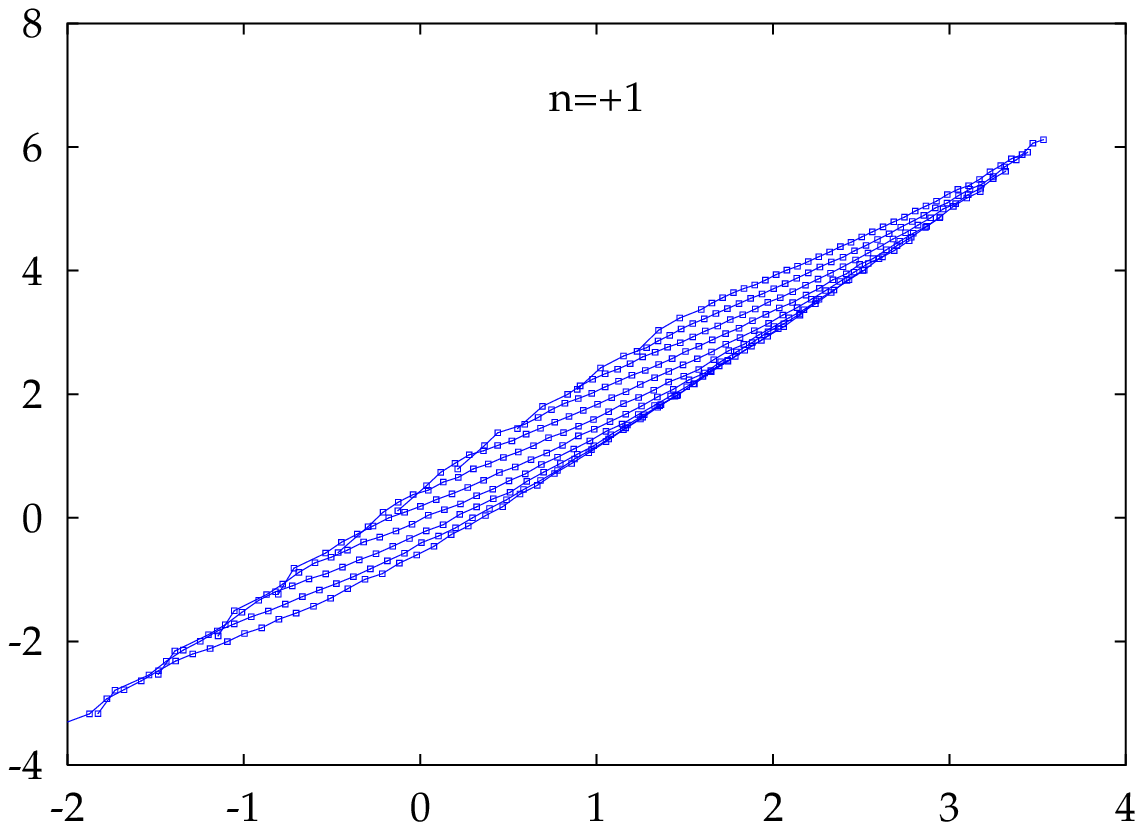}}
  \caption{
    Abscissa: $\log (1+\delta)$. Ordinate: $\log [\ki / \varrho_b
    (\dot{a} L)^2]$. Solid line: theoretical functional dependence,
    Eq.~(\ref{eq:virial}). Plotted only coarsening lengths at the
    latest time, $r_{\rm nl}=R/4$, such that $\sigma(L)>3$. For
    clarity, the scatter bars are omitted; see
    Fig.~\ref{fig:intanom}.}
  \label{fig:intnonlin}
\end{figure}

\begin{figure}
  \resizebox{\hsize}{!}{\includegraphics{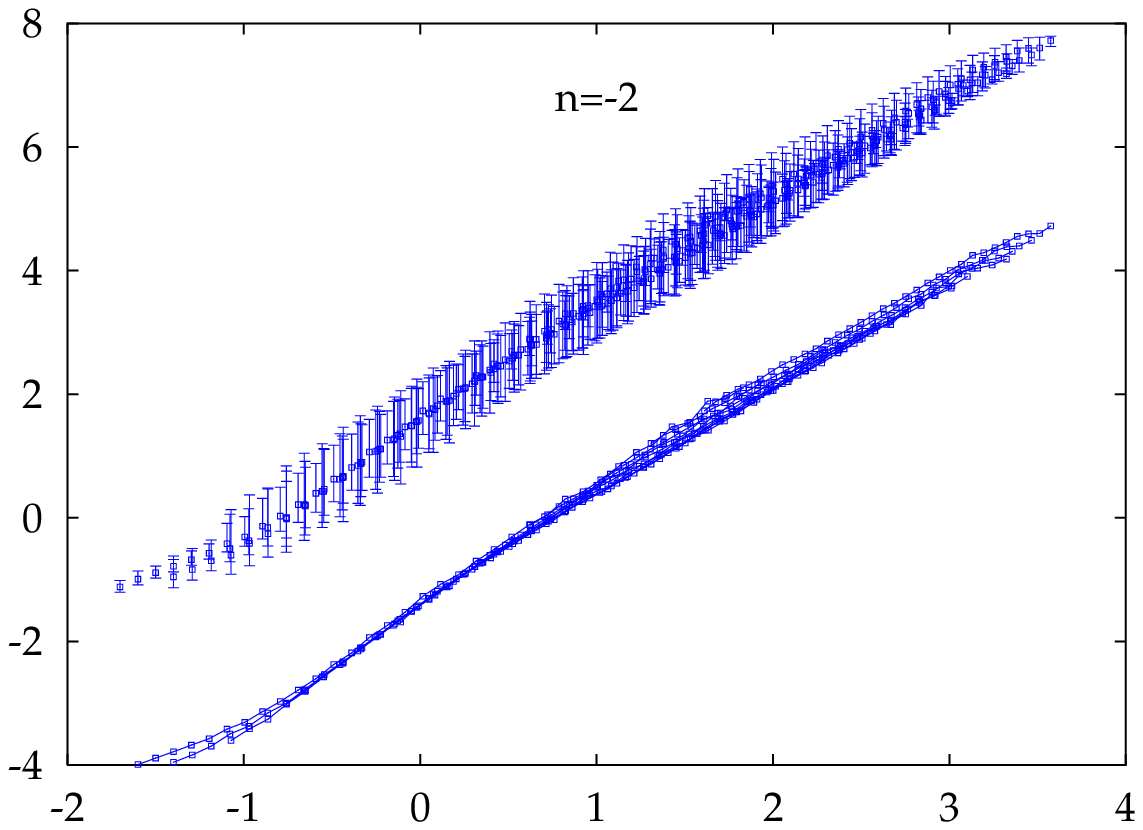}}
  \resizebox{\hsize}{!}{\includegraphics{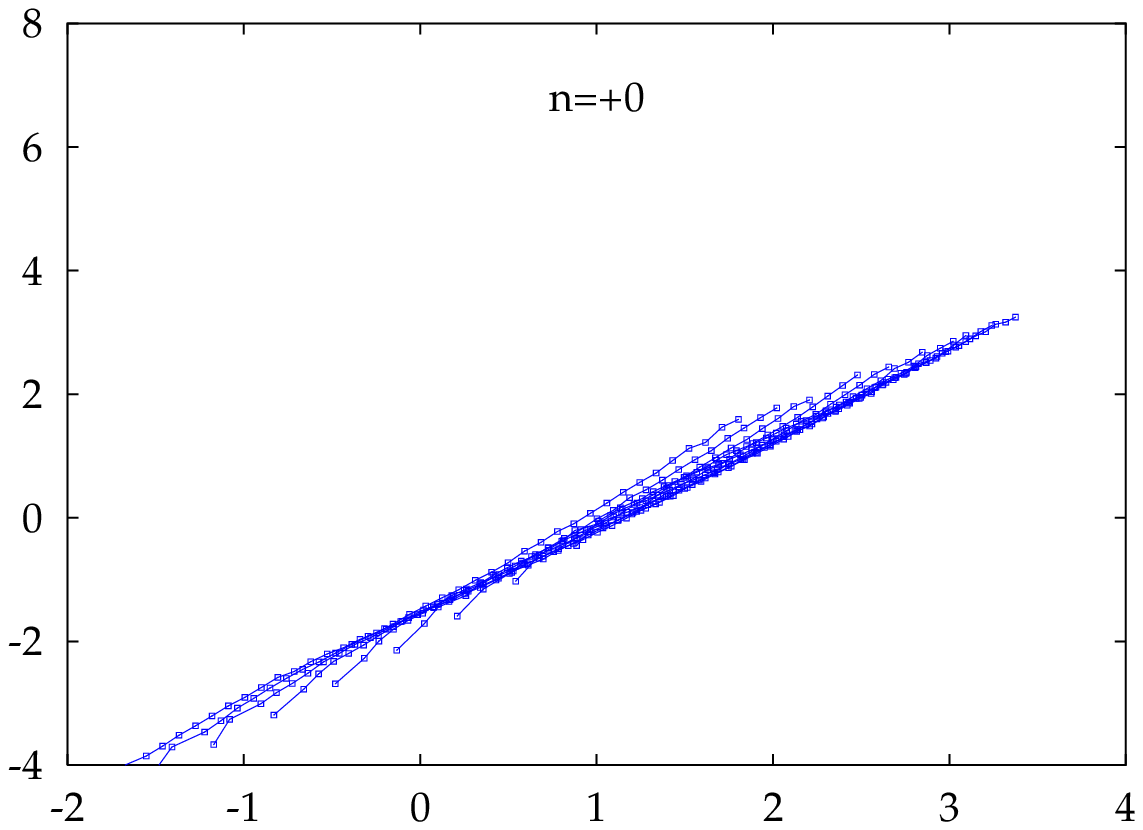}}
  \resizebox{\hsize}{!}{\includegraphics{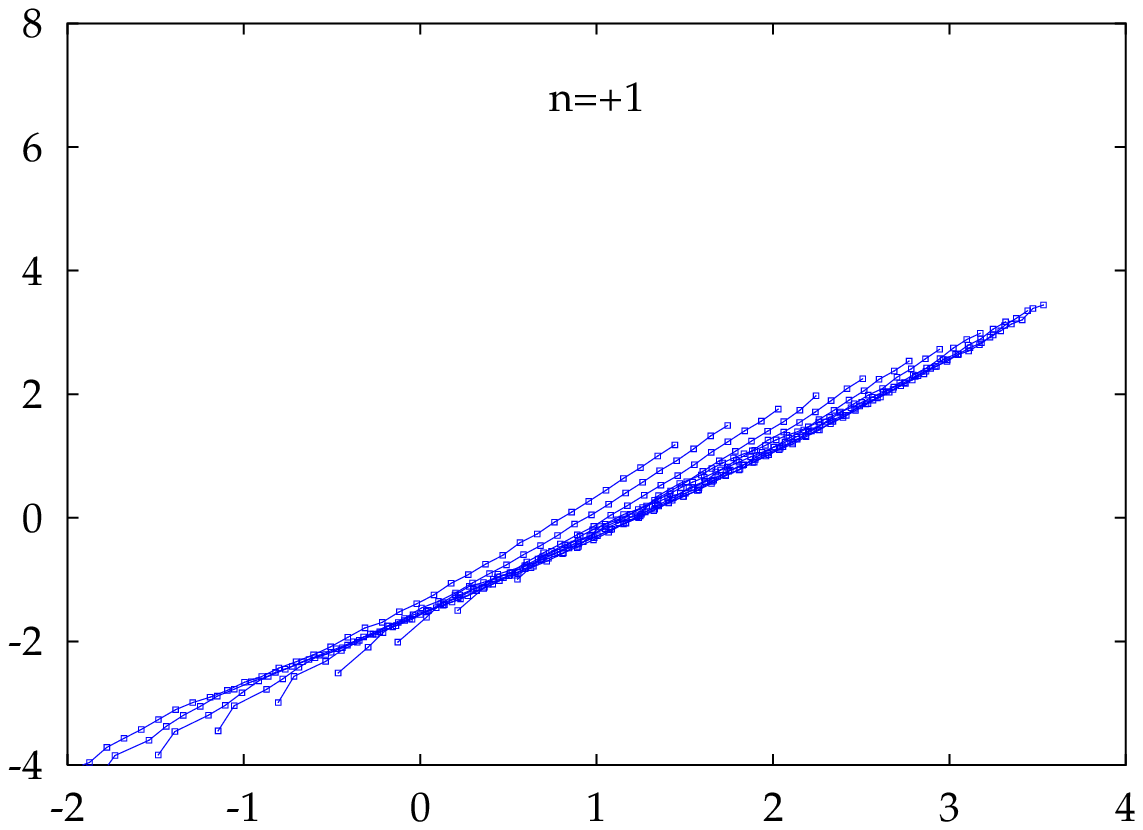}}
  \caption{
    Same as Fig.~\ref{fig:intnonlin}, but now the ordinate is $\log[
    \ki / \varrho_b [\dot{a} L \sigma(L)]^2]$. The solid lines ( $\ki
    \propto \delta^{2-\eta}$) have $\eta=0$ and $\eta (n=-2)
    \approx 0.3$, $\eta (n=0) \approx 0.5$, $\eta (n=+1) \approx 0.6$
    The case $n=-2$ shows also the same data furnished with scatter
    bars and shifted by a factor $10^3$ for clarity. The scatter bars
    of the other two cases are of the same size.}
  \label{fig:intanom}
\end{figure}

Fig.~\ref{fig:intnonlin} shows $\ki$ in the nonlinear
regime at the latest time, $r_{\rm nl}=R/4$.
For each coarsening length, the theoretical functional dependence,
Eq.~(\ref{eq:virial}), is obeyed only at the largest densities, if at all. At
intermediate densities\footnote{The data tend to follow the fit also for very small densities, but mass--resolution effects are large then.}, Fig.~\ref{fig:intanom} shows that the data
can be better fitted by the following polytropic--like behavior:
\begin{equation}
  \label{eq:polytrope}
  \ki \sim \varrho_b (\dot{a} L)^2 \sigma^2(L) \, \delta^{2-\eta} ,
\end{equation}
where the exponent $\eta$ is detectably different from the virial
prediction ($\eta=0$). A fit by eye yields the rough values $\eta
(n=-2) \approx 0.3$, $\eta (n=0) \approx 0.5$, $\eta (n=+1) \approx
0.6$ --- the amount of scatter does not allow a more precise
determination of $\eta$; if the scatter is forgotten, it could even be
that $\eta$ depends slightly on $L$, as seems to be the case when
$n=-2$, Fig.~\ref{fig:intnonlin}.

\begin{figure}
  \resizebox{\hsize}{!}{\includegraphics{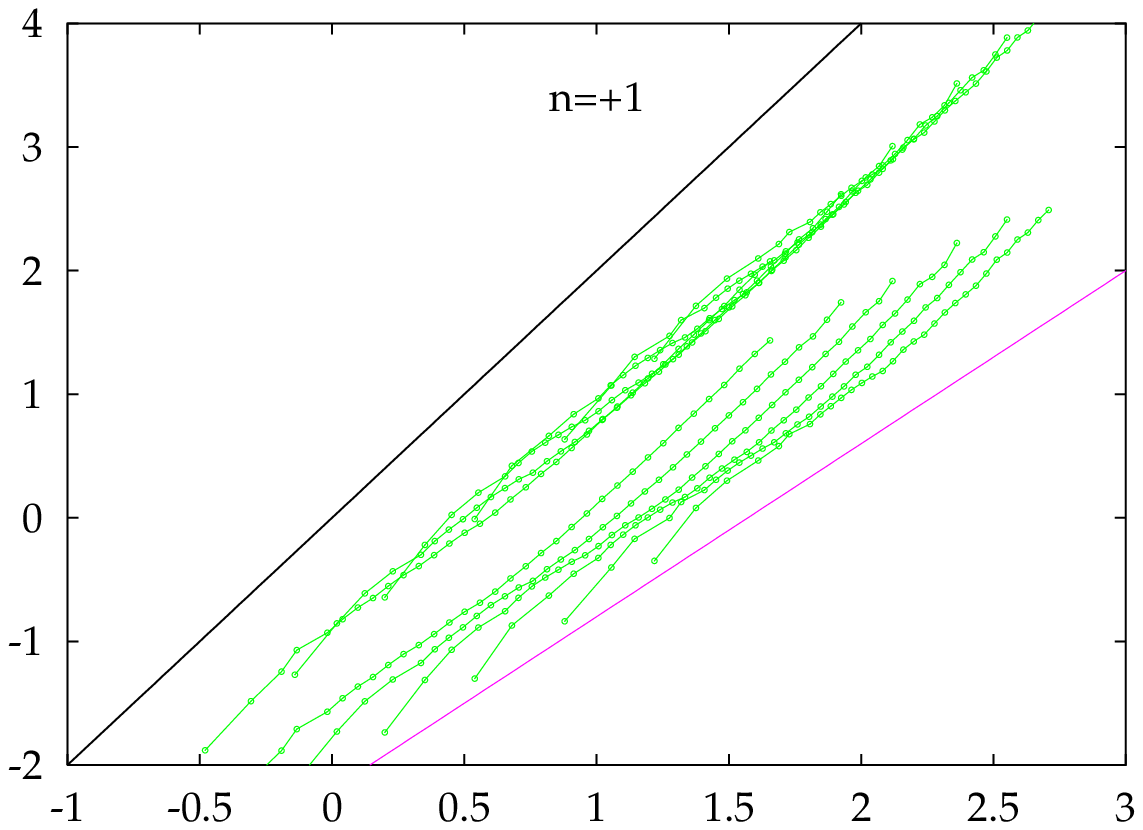}}
  \caption{
    For $n=+1$, $r_{\rm nl}=R/16$, $\sigma(L)>3$: the uppermost data are
    scaled like in Fig.~\ref{fig:intnonlin}, the lowermost data like
    in Fig.~\ref{fig:intanom}. For the other values of $n$, the
    behavior of the data is the same.}
  \label{fig:intanom+1e}
\end{figure}

At the intermediate time, $r_{\rm nl}=R/16$, a similar behavior can be
detected, although the polytropic--like dependence~(\ref{eq:polytrope})
can be discerned only with some difficulty,
Fig.~\ref{fig:intanom+1e}.  Apparently, the dynamical evolution has
not proceeded so far that the intermediate regime with $\eta \neq 0$
can be clearly detected without interference of finite--mass effects:
they show up in this case as an artificial reduction of $\ki$ when
$\delta$ is small enough, i.e., when there are only a few simulation
particles in the coarsening cells, see Fig.~\ref{fig:intanom+1e}.

\subsection{CDM models}
\label{sec:cdm}

\begin{figure}
  \resizebox{\hsize}{!}{\includegraphics{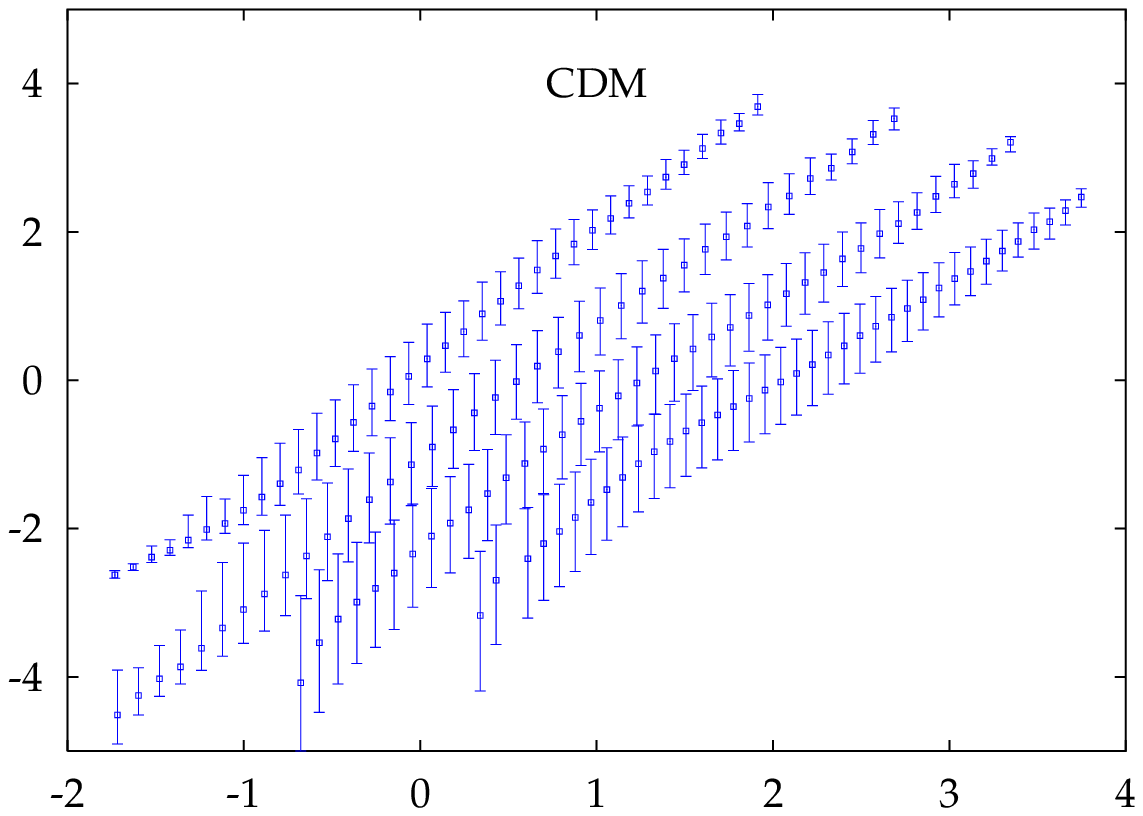}}
  \caption{
    Abscissa: $\log (1+\delta)$. Ordinate: $\log \ki$ (arbitrary
    normalization) for the flat CDM model in a box of sidelength
    $128$~Mpc. For clarity, only some coarsening lengths are plotted
    (decreasing top down). The solid lines ($\ki \propto
    \delta^{2-\eta}$) have $\eta=0$ and $\eta=0.5$.}
  \label{fig:int128flat}
\end{figure}

Regarding the CDM models, we have investigated only the relationship
$\ki (\delta)$ in the nonlinear regime at the present epoch, in order
to assess the possibility of a polytropic--like dependence as in the
self--similar models. Fig~\ref{fig:int128flat} shows the measured $\ki$
as a function of the density for the flat CDM model simulated in a box
of sidelength $128$~Mpc. Now there is no reason to expect an exact
scaling behavior like~(\ref{eq:scaling}) and the plots cannot be made
to collapse on a single function. Nevertheless, a dependence $\ki
\propto \delta^{2-\eta}$ fits well the data for $\delta$ large enough,
with a scale--dependent exponent, $\eta(L)$, which decreases with
decreasing $L$ and ranges between $0$ and $\approx 0.5$ for the
lengths $L$ which we probed 
($L \ga 0.5$ Mpc, see Sec.~\ref{sec:method}).

\begin{figure}
  \resizebox{\hsize}{!}{\includegraphics{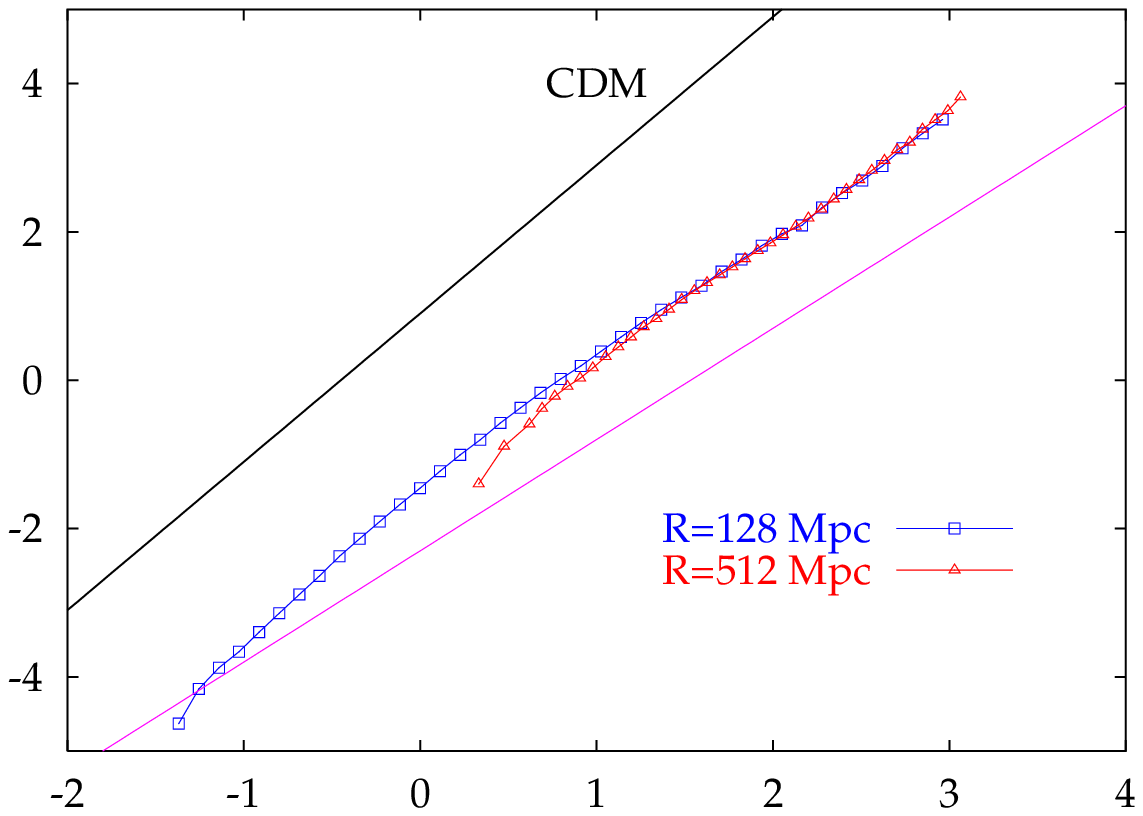}}
  \caption{
    Abscissa: $\log (1+\delta)$. Ordinate: $\log \ki$ (arbitrary
    normalization) for the flat CDM model in boxes of sidelength
    $128$~Mpc and $512$~Mpc, respectively. In the later case, $\ki$
    was multiplied by a factor $10^3$. For clarity, only some
    coarsening lengths are plotted. The solid lines ($\ki \propto
    \delta^{2-\eta}$) have $\eta=0$ and $\eta=0.5$.}
  \label{fig:int128512flat}
\end{figure}

Fig.~\ref{fig:int128512flat} shows the measurements of $\ki$ of the flat
CDM model in two different simulation boxes ($128$ and $512$~Mpc
sidelength, respectively). The worse mass--resolution in the largest
box implies that, for a given $L$ in the highly non-linear regime, the minimum
measurable value of $\delta$ is larger. It also means that the absolute
value of $\ki$ is smaller. The interesting finding is that, if $\ki$
is multiplied by a factor $\approx 10^3$, the plots corresponding to
different simulation boxes but to the same coarsening length superpose
each other. 

The conclusions extracted from the open CDM models are qualitatively
identical to those reached with the flat CDM model, and the numerical
values for the exponent $\eta$ are very similar.

\section{Discussion and conclusions}
\label{sec:discussion}

In the previous Sec.~we have measured the macroscopic and the internal
kinetic energies of cubic cells as a function of time, cell size and
cell mass for different cosmological models. The use of self--similar
models simplifies the task of comparing with theoretical results, when
available. In particular, the scaling relationship~(\ref{eq:scaling})
is useful to assess unphysical dependences on the unavoidable
additional length scales introduced by the simulation procedure,
namely the box sidelength, $R$, and the mean interparticle separation,
$\ell$. It must be noticed that, even when the results obey the
scaling~(\ref{eq:scaling}), this does not imply irrelevance of these
extra length scales: one can conclude at most that $R$ and $\ell$
could enter in the result solely as the combination $R/\ell$, or
equivalently, as $N$, the total particle number. We have not explored
explicitly the influence of such a dependence on $N$. Nevertheless,
from our results we can obtain some hints about how well they
reproduce the limit $N \rightarrow +\infty$.

We first considered the unconstrained averages, $\langle K_{\rm
  int/mac} \rangle$. We find that $\langle K_{\rm mac} \rangle$ for
$n=-2$ suffers from strong finite--size effects in a predictable
manner. The mean internal kinetic energy, $\langle K_{\rm int}
\rangle$, however, is strongly affected by resolution effects, the
more so the larger $n$ is, and its measurement is therefore
unreliable. Next we considered the constrained averages, $\ki$, $\km$.
In general, they are much less affected by resolution effects, which are
``localized'' to very small mass densities or to the earliest time,
being more conspicuous for $n=+1$.

The macroscopic kinetic energy, $\km$, of
the case $n=-2$ depends strongly on $R$, as predicted
theoretically. However, this does not break self--similarity of the
amplitude of density fluctuations, as shown by \citet{JaBe96,JaBe98},
or of $\ki$, as argued in Sec.~\ref{sec:theory} and exemplified by our
results for $\ki$. Relevant for the dynamical evolution of these
physical quantities is not the bulk velocity field, but the {\em
  relative velocity} (that is, the velocity gradient), which does not
suffer this $R$--dependence.

In the cases $n=0,+1$, the theoretical predictions and the
scaling~(\ref{eq:scaling}) are well followed except at the earliest
times and smallest cell sizes, when resolution effects are expected to
be most important. An interesting result is the linear dependence of
$\km$ with $\delta$, Eq.~(\ref{eq:macnonlin01}), with a
proportionality factor which according to Fig.~\ref{fig:macnonlin01}
does not seem to depend sensitively on the spectral index $n$.
\citet{SeSu01} have studied $\km$ in the cases $n=0,+1$ in the quasilinear
regime ($0.2 \la \sigma \la 1.2$), which we have not
addressed at all.

The internal kinetic energy, $\ki$,
is more sensitive to the small scale dynamics than $\km$ is;
correspondingly, resolution effects are found to be more important
than for $\km$. However, only at the earliest time and smallest coarsening cell
sizes for $n=0,+1$ do they render the results unreliable. That's why
the theoretical prediction for $\ki(\delta)$ in the linear regime
could not be tested when $n=0,+1$; when $n=-2$ the reason is that the
linear regime was not really probed, which can be traced back to a
finite--size effect.

In the nonlinear regime, the function $\ki (\delta)$ exhibits an
interesting behavior. The virial prediction, Eq.~(\ref{eq:virial}),
was observed only for $n=0,+1$ asymptotically in the large--$\delta$
end of the curves. Otherwise, a polytropic--like
dependence~(\ref{eq:polytrope}) was found to fit better the data, with an exponent which does
not seem to depend on time or coarsening length, only on the spectral index
$n$. The same polytropic--like dependence is found for CDM models, albeit with a
scale--dependent exponent, confirming the results of an earlier work \citep{Domi03}. The
values of the exponent $\eta$ for the CDM models analyzed here
and by \citet{Domi03} are consistent with each other and with
those of the self--similar models, in spite of the differences in the number of particles
in the simulations (and, compared to \citealt{Domi03}, the simulation
algorithm itself). This suggests that the
polytropic--like relation is not an artifact of the simulations, which
in this respect seem to reproduce acceptably the limit $N \rightarrow
+\infty$. Another hint in this direction is the simple relation which
connects the results of the CDM models in boxes of different size,
Sec.~\ref{sec:cdm}: in the larger box, $R=512$~Mpc, a coarsening cell
of a given mass contains less particles that in the smaller box,
$R=128$~Mpc. Nevertheless, this mass--resolution effect does not alter
the functional dependence $\ki(\delta)$ at all, and can be accounted
for by a scale--independent constant offset.

In the work by \citet{KSS97}, $\ki$
is also measured in CDM simulations (in their notation, $\ki = \varrho
\sigma^2_v /2$). They find a polytropic--like dependence too, but with
a slightly smaller exponent, $\eta \approx 0.1$ \citep[ Eqs.~17-18]{KSS97}. 
We believe this discrepancy to be a consequence of
their simulation having too few particles ($N=32^3$): as a
consequence, they measured the function $\ki(\delta)$ with coarsening
cells having at most 100 particles (in one case); in many cases, the
cells have less than a few tens of particles \citep[ Fig.~3]{KSS97}.
For comparison, the polytropic--like dependence in
Fig.~\ref{fig:int128flat} is detected in coarsening cells containing a
number of particles spanning ranges as wide as $30-7000$ or
$500-30000$ \citep[see also][]{Domi03}.
The work by \citet{KSS97} was motivated by comparison with redshift
surveys. It relied on the cosmic virial theorem and particular
emphasis was put on the dependence with cosmological parameters
($\Omega_m$, $\Omega_\Lambda$). Our results show that departures from
the virial prediction are not small at all, so that the method devised
by \citet{KSS97} must be adjusted. 
More generally, our results warn against a straightforward use of the
cosmic virial theorem to estimate cosmological parameters from
observations without first assessing that the employed observational
data do indeed pertain virialized structures.

In the work by \citet{NOC01}, the cosmic Mach number ($=K_{\rm
  mac}/K_{\rm int}$ in our notation, and $\ki =\varrho \sigma^2 /2$ in
theirs) is measured in a $\Lambda$CDM hydrodynamical simulation, for
three different length scales and as a function of the density. As a
side--result, they also find a polytropic--like dependence for the
velocity dispersion of groups of DM halos and galaxies (with $\eta
\approx 0.5-0.7$) --- the authors do not elaborate much on this result.
One must keep in mind that, compared to our simulations, theirs involves also the baryonic
component and the formation of galaxies, which can affect the velocity
dispersion \citep{TiDo98}.

One can conceive two natural extreme cases of a polytropic--like
dependence: the ``virial'' case, $\eta_{\rm virial}=0$, when velocity
dispersion is fixed by the local mass density, and the ``isothermal''
case, $\eta_{\rm isothermal}=1$, when velocity dispersion is fixed by
an external cause, e.g.~tidal forces, free flow,... The values of $\eta$ that we
measure invariably fall between $0$ and $1$; the corresponding
relations $\ki(\delta)$ can be arguably understood as the outcome of
the competition of the two effects (``local virialization vs.~global
thermalization''), whose relative strength varies with the spectral
index $n$ and the cell size and mass. However, an elaborated theory is
required to lend support to this explanation, 
the ultimate goal being the ``postdiction'' of the
relation~(\ref{eq:polytrope}).
The value $\eta=1/3$ was derived theoretically by \citet{BuDo98}, but
we think this is irrelevant to our results, since 
certain restrictive assumptions were made (vanishingly small and isotropic
velocity dispersion, approximately shear--free velocity field $\bu$),
which are unlikely to hold in the regime where we find the
polytropic--like dependence.
%
The results from the simulations cannot be explained by any theory
%
whose starting point is the usual thermodynamical theory or, more
generally, the (grand--)canonical ensemble of statistical mechanics
\citep{SaFa96,HoPe96,VSC98}, since in that framework the
kinetic energy is an extensive variable: $\ki = T \varrho$ ($T$ is the
kinetic temperature) and $\eta=1$.
As a side--remark, we notice that \citet{SCII90} compute the velocity
distribution allegedly in the framework of thermodynamics: but they
use contradictory arguments
and obtain instead that the kinetic energy scales like
$\varrho^2$ (that is, $\eta=0$), and a velocity distribution different
from the Maxwellian one characteristic of thermal equilibrium and
which should follow from the (grand--)canonical ensemble probability.
%

The discovered relationship $\ki(\varrho)$ is useful for an improved
model of structure formation by gravitational instability
\citep{BuDo98}: the dust model (pressureless fluid) is added a term
proportional to the gradient of $\ki$ (a kinetic pressure), in order
to account for the reaction of the dynamically generated velocity
dispersion on the evolution. 
The evolution equation for the velocity field $\bu (\bx, t)$ then
reads
\begin{equation}
  \label{eq:euler}
  \frac{\partial \bu}{\partial t} = - \frac{\dot{a}}{a} \bu 
  - \frac{1}{a} (\bu  \cdot \nabla) \bu + {\bf w}
  - \frac{1}{a  \varrho} \nabla \ki ,
\end{equation}
where the peculiar gravitational acceleration ${\bf w}$ is given by
Poisson's equation.
Further theoretical studies of this model
\citep{AdBu99,BDP99,MTM99,MoTa01,TSMM02} work with a $\ki$ which
depends only on density, e.g.~a pure polytropic--like dependence with
values of the exponent $\eta$ in concordance with our measurements.
Our results show that the functional dependence of $\ki$ on $\varrho$
is somewhat more complicated than purely polytropic and changes with
time and coarsening length.
Nevertheless, since the term $- \nabla \ki$ is a pressure, it opposes
compression in collapsing regions.
More can be learned about the behavior of this term when some
simplifications are introduced \citep{BuDo98,BDP99}: one assumes that
the evolution follows the dust model prediction in the form of the
Zel'dovich approximation (basically that $\bu \propto {\bf w}$)
``almost everywhere'', i.e., except near potential density
singularities where the effect of $-\nabla \ki$ becomes relevant. One
can then apply boundary layer theory to show \citep{Domi00} that this
term does behave ``adhesively'' and indeed succeeds in preventing the
formation of a singularity, provided $\ki$ is a function of $\varrho$
and $\ki(\varrho)/\varrho$ is a growing function of $\varrho$ (meaning
$\eta <1$ for a polytropic--like dependence). This behavior is robust
against the observed time--dependence in the relation $\ki(\varrho)$,
being much slower than the time--scale of collapse. One can conclude
that the dependence $\ki(\varrho; t)$ measured in N--body simulations
leads to the same qualitative ``adhesive'' behavior as the simpler
dependences addressed theoretically in the literature.

When it comes to inserting our results in the theoretical
model~(\ref{eq:euler}),
there are some issues which we have not addressed but may be relevant
to a better understanding of the model.
First, it must be noticed that the
average relationship $\ki(\varrho)$ does not mean in principle a
one--to--one dependence between $K_{\rm int}$ and $\varrho$; on the
contrary, the data scatter around the average dependence,
Fig.~\ref{fig:intanom}.  
In fact, the derivation of Eq.~(\ref{eq:euler}) yields in reality a
term $-\nabla K_{\rm int}$ \citep{BuDo98}:
the influence of the scatter on the model outputs should be quantified
and, if proven relevant, incorporated in the model, e.g.~as a noisy
source \citep{BDP99}. 
Another issue of possible concern 
is the amount of velocity dispersion
in the coarsening cells associated to ``bound structures'' (as opposed
to the amount associated to particle flow between neighboring cells);
in this context, it would also be interesting to assess the
contribution to velocity dispersion from ``ordered motion'', e.g.~due
to a net angular momentum.

In conclusion, we have studied the density dependence of the
macroscopic and internal kinetic energies in coarsening cells. We
could identify the influence of finite--size and resolution effects on
the measured physical quantities. When these effects were irrelevant,
we could confirm some of the theoretical asymptotic predictions.
Finally, we found that in an intermediate range of densities, the
velocity dispersion scales as a power of the mass density, with an
exponent different from the virial prediction.

\begin{acknowledgements}
  
  A.D. acknowledges support of the ``Sonder\-forschungs\-bereich SFB
  375 f\"ur Astro--Teilchenphysik der Deutschen
  Forschungsgemeinschaft''.  A.L.M. acknowledges support of US NSF
  through grant AST0070702 and the National Center for Supercomputing
  Applications (Urbana, Illinois, USA).

\end{acknowledgements}

\newpage


\appendix

\section{Estimates with the linear solution}
\label{ap:math}

In this Appendix we collect the mathematical calculations which lead
to Eqs.~(\ref{eq:mac2}-\ref{eq:intlin2}). The main idea is that
$\ki$ and $\km$ are determined by the dominant
contribution of modes in the linear regime when $L \gg r_{\rm nl}$, so that
they can be estimated by inserting the linear solution in the
definitions~(\ref{eq:fields}).  These definitions can be rewritten as
follows:
\begin{displaymath}
\varrho (\bx, t) = \int {{\rm d}\by \over L^3} \; W \left( {\bx -
    \by \over L} \right) \varrho_{mic} (\by, t) ,
\end{displaymath}
\begin{displaymath}
  \varrho \bu (\bx, t) = \int {{\rm d}\by \over L^3} \; W \left( {\bx
      - \by \over L} \right) \varrho_{mic} \bu_{mic} (\by, t) ,
\end{displaymath}
\begin{displaymath}
  K_{\rm int} (\bx, t) = \frac{1}{2} \int {{\rm d}\by \over L^3} \; 
  W \left( {\bx - \by \over L} \right) 
  \varrho_{mic} \bu_{mic}^2 (\by, t) - \frac{1}{2} \varrho \bu^2 (\bx, t),
\end{displaymath}
in terms of the formal microscopic fields\footnote{In the mathematical
  manipulations in this Appendix, one can assume that the Dirac delta
  is regularized by a short--distance cutoff which is set to zero in
  the final results. In this way, the field $\bu_{mic} (\bx)$ is
  defined, and not only the product $\varrho_{mic} \bu_{mic} (\bx)$.}
\begin{displaymath}
\varrho_{mic} (\bx, t) := \frac{m}{a(t)^3} \sum_\alpha \delta^{(3)}
(\bx - \bx_\alpha (t)) ,
\end{displaymath}
\begin{displaymath}
  \varrho_{mic} \bu_{mic} (\bx, t) := \frac{m}{a(t)^3} \sum_\alpha
  \bu_\alpha (t) \delta^{(3)} (\bx - \bx_\alpha (t)) .
\end{displaymath}

We introduce the Fourier transform of any spatial field $\phi({\bf
  x})$, denoted by a tilde and defined as
\begin{displaymath}
  \tilde\phi({\bf k}) := \int \!\! {\rm d} {\bf x} \;
  {\rm e}^{i \bk \cdot \bx} \, \phi({\bf x}) .
\end{displaymath}
The velocity then reads
\begin{displaymath}
  \varrho \bu (\bx, t) = \int \frac{{\rm d} \bk \, {\rm d} \bq}{(2 \pi)^6} 
  \, {\rm e}^{- i (\bk+\bq) \cdot \bx} \, \tilde\varrho_{mic} (\bk, t) 
  \tilde\bu_{mic} (\bq, t) \tilde{W} (L (\bk+\bq)) ,
\end{displaymath}
and the internal kinetic energy
\begin{equation}
  \label{eq:kintfourier}
  K_{\rm int} (\bx, t) = \frac{1}{2} 
  \int \frac{{\rm d} \bk \, {\rm d} \bq \, {\rm d} \bp}{(2 \pi)^9} 
  \, {\rm e}^{- i (\bk+\bq+\bp) \cdot \bx} \, \tilde\varrho_{mic} (\bk, t) \times
\end{equation}
\begin{displaymath}
   \times [ \tilde\bu_{mic} (\bq, t) \cdot \tilde\bu_{mic} (\bp, t) 
   \tilde{W} (L (\bk+\bq+\bp)) - \tilde\bu (\bq, t) \cdot \tilde\bu (\bp, t) \tilde{W} (L \bk) ] .
 \end{displaymath}
Even if $L \ll r_{\rm nl}$, these expressions contain contributions
from nonlinear modes ($r_{\rm nl} k \gg 1$). The hypothesis
(Sec.~\ref{sec:theory}) is that these contributions are nevertheless
negligible compared to those from the linear modes ($r_{\rm nl} k \ll
1$). This implies $\tilde\bu (\bk) \approx \tilde\bu_{mic} (\bk)
\tilde{W} (L \bk)$ in the linear regime, and the growing linear solution in an Einstein--de
Sitter background yields in turn
\begin{displaymath}
\tilde\bu_{mic} (\bk, t) \approx - \dot{a} \frac{i \bk}{k^2} 
\tilde\delta_{mic} (\bk, t) , \qquad\qquad (r_{\rm nl} k \ll 1)
\end{displaymath}
with $\delta_{mic} = (\varrho_{mic}/\varrho_b) - 1$. Inserting these
linear relations in the definition~(\ref{eq:kintfourier}) we obtain,
to lowest order in the inhomogeneities,
\begin{displaymath}
  K_{\rm int} (\bx, t) \approx \frac{1}{2} \varrho_b \dot{a}^2
  \int \frac{{\rm d} \bk \, {\rm d} \bq}{(2 \pi)^6} 
  \, {\rm e}^{- i (\bk+\bq) \cdot \bx} \, \frac{\bk \cdot \bq}{k^2 q^2} \times
\end{displaymath}
\begin{displaymath}
\quad \times
  [\tilde{W} (L \bk) \tilde{W} (L \bq) - \tilde{W} (L (\bk+\bq))] \, 
  \tilde\delta_{mic} (\bk, t) \tilde\delta_{mic} (\bq, t) .
\end{displaymath}

For Gaussian initial conditions, the average of
Eq.~(\ref{eq:eqstate}) can be easily computed using standard
techniques for the Gaussian functional integrals \citep[e.g.~][]{Zinn96}:
\begin{displaymath}
\langle \delta^{(3)} (\varrho({\bf x}, t; L)-\varrho) \rangle = 
\frac{1}{\sqrt{2 \pi} \varrho_b (t) \sigma(L,t)}
\exp \left(-\frac{\delta^2}{2 \sigma^2 (L, t)}\right) , 
\end{displaymath}
\begin{displaymath}
\frac{\langle \tilde\delta_{mic} (\bk, t) \tilde\delta_{mic} (\bq, t) 
\delta^{(3)} (\varrho({\bf x}, t; L)-\varrho) \rangle}
{\langle \delta^{(3)} (\varrho({\bf x}, t; L)-\varrho) \rangle} =
(2 \pi)^3 P(k,t) \delta^{(3)} (\bk+\bq) - 
\end{displaymath}
\begin{displaymath}
\qquad - \frac{1}{\sigma^2(L, t)} \left(1 - \frac{\delta^2}{\sigma^2(L, t)} \right) 
P(k, t) P(q, t) \tilde{W}(L \bk) \tilde{W}(L \bq) {\rm e}^{i (\bk+\bq) \cdot \bx} .
\end{displaymath}
With these expressions, one recovers the result~(\ref{eq:intlin2})
with the coefficients given by 
\begin{equation}
  \label{eq:meankin}
  \langle K_{\rm int} \rangle (L, t) = \frac{1}{2} \varrho_b \dot{a}^2 
  \int \frac{{\rm d} \bk}{(2 \pi)^3} \frac{P(k)}{k^2} [1 - |\tilde{W}(L \bk)|^2] ,
\end{equation}
\begin{equation}
  \label{eq:Bcoeff}
  B  (L, t) = \frac{\varrho_b \dot{a}^2}{2 \sigma^2 \langle K_{\rm int} \rangle (L, t)} 
  \int \frac{{\rm d} \bk \, {\rm d} \bq}{(2 \pi)^6} \frac{\bk \cdot \bq}{k^2 q^2} \times
\end{equation}
\begin{displaymath}  
\qquad\qquad \times P(k, t) P(q, t) \tilde{W}(L \bk) \tilde{W}(L \bq) \tilde{W}(L (\bk+\bq)) .
\end{displaymath}
These integrals are IR--convergent provided $P(k \rightarrow 0) \sim
k^n$, $-3<n$, since $\tilde{W} (\bk \rightarrow {\bf 0}) \sim 1 +
o(k^2)$ ($\leftrightarrow$ the window function is normalized to unity
and decays fast enough at large distances). On the other hand, there
is an implicit ultraviolet (UV) cutoff, $k, q \leq r_{\rm nl}^{-1}$, because these
expressions have been derived for the linear regime. 
The limit ``$r_{\rm nl} \rightarrow 0$'' of the coefficient $B$ is finite, 
since we assume that
$\tilde{W} (\bk \rightarrow \infty) \rightarrow 0$ fast enough
($\leftrightarrow$ the window is smooth enough: the decay is
exponential for a Gaussian window, algebraic ($\sim k^{-2} \cos k$)
for a spherical top--hat window). $\langle K_{\rm int} \rangle$ is however
UV--convergent only if $P(k \rightarrow \infty) \sim k^n$, $n<-1$, 
indicating that it is determined by nonlinear modes when $n>-1$.

A similar reasoning can be repeated for the macroscopic kinetic
energy: inserting the linear relationships in the definition $K_{\rm mac}
= (1/2) \varrho \bu^2$ one gets

\begin{displaymath}
  \km \approx \frac{1}{2} \varrho_b \dot{a}^2
  \int \frac{{\rm d} \bk}{(2 \pi)^3} \frac{P(k)}{k^2} |\tilde{W}(L \bk)|^2
  = \langle K_{\rm mac} \rangle ,
\end{displaymath}
which is IR--divergent in the range of spectral indices $-3<n<-1$.

\end{document}